\documentclass[letterpaper,journal]{IEEEtran}
\usepackage{amsmath,amsfonts}
\usepackage{algorithm}
\usepackage{graphicx}
\usepackage{array}
\usepackage[caption=false,font=normalsize]{subfig}
\usepackage{textcomp}
\usepackage{stfloats}
\usepackage{url}
\usepackage{verbatim}
\usepackage{graphicx}
\usepackage{cite}

\usepackage{booktabs}   
\usepackage{bm}
\usepackage{multirow}
\usepackage{upgreek}
\usepackage{siunitx} % 用于数字格式化
\usepackage{array} % 提供额外的表格功能
\usepackage[T1]{fontenc}
\usepackage{courier}
\usepackage{float}
\usepackage{threeparttable}
\usepackage{url}
\usepackage{hyperref}
\usepackage[T1]{fontenc}
\usepackage{makecell}
\usepackage{stmaryrd}
\usepackage{xcolor}
\definecolor{midgray}{RGB}{128,128,128}
\definecolor{myblue}{RGB}{189,215,238}
\definecolor{mypink}{RGB}{246,194,194}
\definecolor{mygreen}{RGB}{208,229,203}
\definecolor{myyellow}{RGB}{255,230,153}
\definecolor{mygold}{RGB}{255,217,102}
\definecolor{myorange}{RGB}{246,142,16}
\definecolor{pinblue}{RGB}{32,215,253}
\definecolor{macroyellow}{RGB}{221,218,175}
\definecolor{commentcolor}{RGB}{20,40,210}
\usepackage[commentColor=commentcolor]{algpseudocodex} % added for pseudo code
\begin{document}

\title{$\text{Re}^{\text{2}}$MaP: \underline{Ma}cro \underline{P}lacement by \underline{Re}cursively Prototyping and \\
Packing Tree-based \underline{Re}locating}

\author{Yunqi Shi*, Xi Lin*, Zhiang Wang,~\IEEEmembership{Member,~IEEE,} Siyuan Xu, Shixiong Kai, Yao Lai, Chengrui Gao, Ke Xue, Mingxuan Yuan,~\IEEEmembership{Member,~IEEE,} Chao Qian,~\IEEEmembership{Senior Member,~IEEE,} and Zhi-Hua Zhou,~\IEEEmembership{Fellow,~IEEE}
        % <-this % stops a space
\thanks{Manuscript received April 19, 2021; revised August 16, 2021. (\textit{Corresponding author: Chao Qian.})}
\thanks{Yunqi Shi, Xi Lin, Chengrui Gao, Ke Xue, Chao Qian and Zhi-Hua Zhou are with the National Key Laboratory for Novel Software Technology and the School of Artificial Intelligence, Nanjing University, Nanjing, China (email: qianc@lamda.nju.edu.cn).}
\thanks{Zhiang Wang is with the College of Integrated Circuits and Nano-Micro Electronics, Fudan University, Shanghai, China.}
\thanks{Siyuan Xu is with Huawei Noah's Ark Lab, Shenzhen, China.}
\thanks{Shixiong Kai is with Huawei Noah's Ark Lab, Beijing, China.}
\thanks{Yao Lai is with the School of Computing and Data Science, the University of Hong Kong, Hong Kong, SAR.}
\thanks{Mingxuan Yuan is with Huawei Noah's Ark Lab, Hong Kong, SAR.}
\thanks{*Equal contribution.}}

\markboth{Journal of \LaTeX\ Class Files,~Vol.~14, No.~8, August~2021}%
{Shell \MakeLowercase{\textit{et al.}}: A Sample Article Using IEEEtran.cls for IEEE Journals}

\IEEEpubid{0000--0000/00\$00.00~\copyright~2021 IEEE}

\maketitle

\begin{abstract}
Macro placement plays a vital role in very large-scale integration (VLSI) backend design, since macros typically occupy large areas, and poor placement may complicate power distribution, degrade total wirelength and timing, and cause severe routing congestion. However, this stage still relies highly on human experts, and few automatic placers can produce reasonable layouts that meet industrial requirements. Thus, we introduce the Re$^{\text{2}}$MaP method, which generates expert-quality macro placements through recursively prototyping and packing tree-based relocating. We first perform multi-level macro grouping and PPA-aware cell clustering to produce a unified connection matrix that captures both wirelength and dataflow among macros and clusters. Next, we use DREAMPlace to build a mixed-size placement prototype and obtain reference positions for each macro and cluster. Based on this prototype, we introduce ABPlace, an angle-based analytical method that optimizes macro positions on an ellipse to distribute macros uniformly near chip periphery, while optimizing wirelength and dataflow. A packing tree-based relocating procedure is then designed to jointly adjust the locations of macro groups
and the macros within each group, by optimizing an expertise-inspired cost function that captures various design constraints through evolutionary search. Re$^{\text{2}}$MaP repeats the above process: Only a subset of macro groups are positioned in each iteration, and the remaining macros are deferred to the next iteration to improve the prototype’s accuracy. Using a well-established backend flow with sufficient timing optimizations, Re$^{\text{2}}$MaP achieves up to 22.22\% (average 10.26\%) improvement in worst negative slack (WNS) and up to 97.91\% (average 33.97\%) improvement in total negative slack (TNS) compared to the state-of-the-art academic placer Hier-RTLMP~\cite{kahng2023hier}. It also ranks higher on WNS, TNS, power, design rule check (DRC) violations, and runtime than the conference version ReMaP~\cite{shi2025remap}, across seven tested cases. Our code is available at \url{https://github.com/lamda-bbo/Re2MaP}.

\end{abstract}


\section{Introduction}

Macro placement determines the locations of macros within a fixed outline, which is a key step in the floorplanning stage and is critical to the entire backend flow~\cite{mirhoseini2021graph,shi2025open3dbench}. Macro placement directly sets the placeable area for standard cells~\cite{lin2019novel}, where irregular macro layouts create dead space and notch regions that analytical cell placers cannot utilize effectively, which will degrade global placement wirelength quality~\cite{pu2024incremacro}, leading to inferior timing and power performance~\cite{shi2025timing}. In addition, when small gaps between macros are filled with standard cells~\cite{kahng2023hier}, the power delivery network (PDN) often cannot reach these cells reliably, causing serious power integrity (PI) issues~\cite{lin2019regularity}. Macro placement also affects timing optimization by determining whether there is sufficient space for buffer insertion and gate sizing~\cite{kahng2025recursive}. It further impacts routing convergence, because macros typically reserve multiple metal layers for internal routing~\cite{pu2024incremacro}.

Despite decades of research and mature commercial placers~\cite{gigaplace}, the final macro layout in industrial design flows is still largely decided by human experts. The reason is twofold: macro placement has outsized impact on global interconnect and congestion patterns; more critically, it is governed by a set of stringent, design-specific constraints that vary across projects, such as hierarchy consistency, datapath orientation, macro grouping and tiling, overall placement regularity, periphery placement, and I/O region keepout. Because these constraints are diverse, discrete, and strongly coupled, existing placers struggle to model and satisfy them reliably across a broad range of designs, preventing full automation. 

Balancing scalability and constraint expressiveness, current approaches fall into three broad categories: metaheuristic-based placers, analytical mixed-size placers, and learning-based placers.

\IEEEpubidadjcol

\textbf{Metaheuristic-based methods}~\cite{ousterhout1984corner,murata1996vlsi,hong2000corner,wu2004placement,yan2009handling} are widely used in industrial macro placement flows because they are flexible, robust, and easy to implement~\cite{mirhoseini2021graph}. They typically use a placement representation (i.e., a data structure) and iteratively improve solutions with stochastic search, represented by simulated annealing (SA). Early work focused on efficient representations for rectangular packing, including corner stitching~\cite{ousterhout1984corner}, sequence pair~\cite{murata1996vlsi}, corner block list~\cite{hong2000corner}, B*-tree~\cite{wu2004placement}, and slicing tree~\cite{yan2009handling}. This setting is classified into ``floorplanning'' and aims to minimize wirelength and total bounding box area. Such packing representations are then adapted for modern fixed-outline macro placement for mixed-size designs (e.g., MP-trees~\cite{chen2007mptrees} built on B*-trees for macro corner packing), since expert practices favor regularly packed solutions. Today’s evaluation criteria include wirelength, routability, and macro displacement relative to mixed-size prototyping. However, these metrics alone may not capture communication locality at the register transfer level (RTL); consequently, dataflow has recently emerged as an optimization target~\cite{vidal2019rtl,lin2019novel,lin2021dataflow}, offering RTL-level guidance for macro placement via dataflow affinity between macros and cell clusters~\cite{zhao2024standard}. In addition, leading placers account for design hierarchy, I/O region keepout, macro grouping, and notch avoidance~\cite{kahng2022rtl,kahng2023hier}, to satisfy the design constraints and achieve expert-quality layouts.

\textbf{Analytical methods}~\cite{lin2019dreamplace, chen2023stronger, cheng2018replace, lin2019routability} optimize wirelength and placement density to place macros and standard cells simultaneously. These methods are efficient and provide a global view for wirelength optimization. DAPA~\cite{lin2021dapa} and DG-RePlAce~\cite{kahng2024dg} incorporate dataflow constraints into global placement to capture critical datapaths and optimize them accordingly. IncreMacro~\cite{pu2024incremacro} identifies macros that drift away from the periphery and introduces a smoothed cost as a nonlinear optimization metric. RegPlace~\cite{jiang2025regplace} targets regularity-driven mixed-size placement for deep neural network (DNN) accelerators, enabling joint optimization of macros and process element (PE) standard cells.

\textbf{Learning-based methods}~\cite{lin2025datemp,mirhoseini2021graph,lai2022maskplace,lai2023chipformer,geng2024reinforcement,xue2024reinforcement,tashdid2025beyondppa} have gained attention following Google’s AlphaChip~\cite{mirhoseini2021graph,goldie2024addendum}. Most approaches use reinforcement learning (RL) for macro placement but face challenges in reward design. Providing a terminal reward only after all macros are placed~\cite{mirhoseini2021graph, cheng2021joint} causes sparse feedback, while using incremental wirelength as a dense reward~\cite{lai2022maskplace, lai2023chipformer, shi2024macro, geng2024reinforcement} often gathers macros together and degrades power, performance, and area (PPA) performance. MaskRegulate~\cite{xue2024reinforcement} explicitly promotes periphery placement to improve timing and reduce routing congestion. BeyondPPA~\cite{tashdid2025beyondppa} emphasizes post-route reliability and is the first learning-based method to encode I/O region keepout, notch constraints, and macro alignment into both features and optimization.

\textbf{Drawbacks of existing work} are evident. Metaheuristic-based methods cannot effectively use prototyping, since minimizing macro displacement often conflicts with other design constraints such as periphery placement and I/O region keepout. In addition, most of them perform prototyping only once~\cite{chen2007mptrees,lin2022novel}, which is insufficient, because the estimates for the remaining macros and cells become inaccurate after a macro moves, requiring recomputation. For analytical placers, they cannot fully encode complex physical constraints into a numeric optimization objective, and thus cannot directly produce expert-quality layouts. They typically require heuristics for legalization or refinement~\cite{chen2023stronger,pu2024incremacro}. For learning-based methods, we currently do not have access to adequate training data, and the transferability to real industrial designs still requires further validation.

\textbf{Our proposed solution is Re$^{\text{2}}$MaP}, a multi-stage macro placement method that combines recursive prototyping, angle-based analytical macro optimization on an ellipse, and packing tree-based macro relocating. Our approach leverages the global wirelength awareness of analytical methods through recursively applied mixed-size prototyping, and the robustness of metaheuristics by relocating macros using a tree representation with evolutionary search~\cite{zhou2019evolutionary}.

\begin{figure}[t]
\centering
\includegraphics[width=0.46\textwidth]{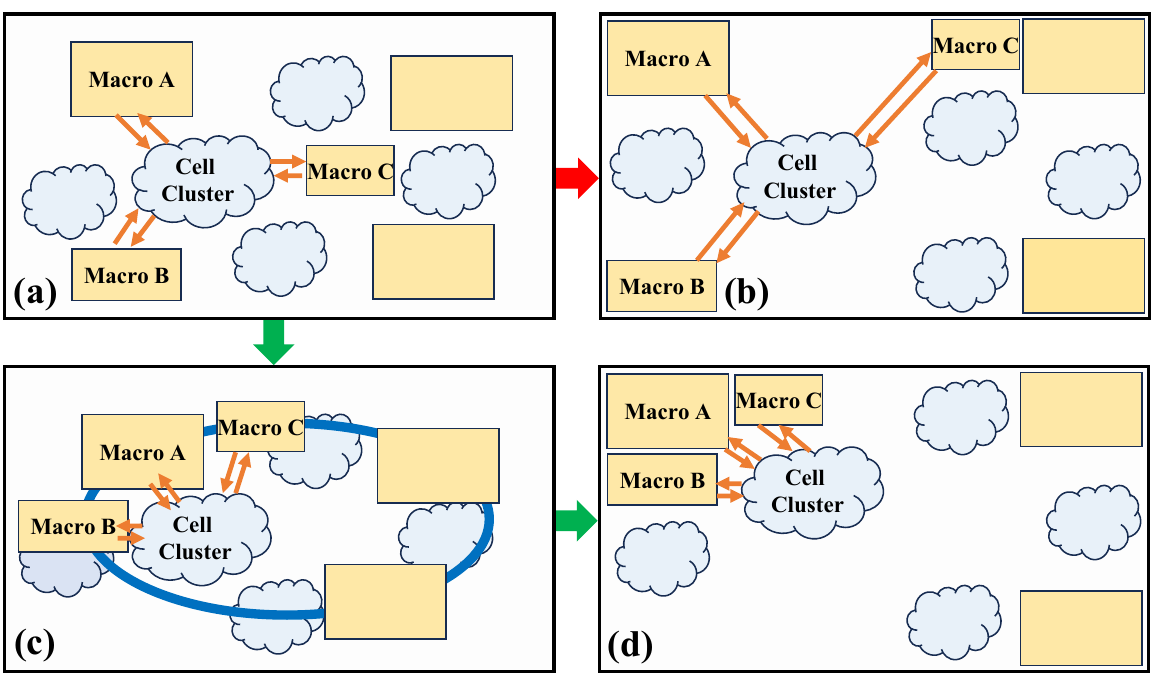}
\caption{(a) Prototype of mixed-size placement. (b) Placement result following the direct application of a packing-based heuristic that minimizes macro displacement. (c) Optimization of macros on an ellipse by ABPlace. (d) Placement result by packing tree-based relocating. The \textcolor{myorange}{orange} arrows represent wirelength and dataflow connections.}
\vspace{-0.5\baselineskip}
\label{fig:intro}
\end{figure}

The flow is illustrated in Fig.~\ref{fig:intro}. Subfigure~(a) shows a mixed-size prototyping result. Moving from subfigure~(a) to subfigure~(b), the macro displacement is minimized using a corner-packing data structure. However, the total wirelength becomes worse because macros A, B, and C are all connected to the same cell cluster but end up separated. Subfigures~(a), (c), and (d) depict our optimization procedure. First, we construct an ellipse to distribute macros near the chip periphery. Next, we perform angle-based analytical optimization on the ellipse to minimize total dataflow and wirelength cost. Unlike traditional metaheuristics that optimize a weighted sum of all objectives at once, we focus on wirelength and dataflow at this stage; other conflicting objectives are deferred to the next stage. Finally, macros are assigned to corners using a packing tree representation, and the design constraints including macro grouping, periphery placement, I/O region keepout, and notch avoidance are enforced via an evolutionary search guided by a joint cost function. Steps (a), (c), and (d) are applied recursively. At each iteration, only a subset of macro groups are fixed to preserve prototyping fidelity.

Our contributions are summarized as follows:
\begin{itemize}
\item We develop Re$^{\text{2}}$MaP, a macro placement method that delivers expert-quality layouts through a recursive process of prototyping and relocating. The algorithm naturally handles pre-placed blocks by iteratively placing and fixing macros in stages. We have open-sourced the code at \url{https://github.com/lamda-bbo/Re2MaP}.
\item We propose a unified macro grouping and PPA-aware cell clustering method, and extract a macro–macro and macro–cluster connection matrix that captures both wirelength and dataflow.
\item We introduce ABPlace, a novel angle-based analytical method that arranges and optimizes macros along an ellipse. This approach considers both wirelength and dataflow between macros and macro–cell clusters, and promotes a uniform distribution of macros near the chip periphery. 
\item We propose an efficient packing tree-based macro relocating method, which tries to pack macro groups to four corners of the chip. We carefully design a multi-stage optimization method including early assessment and correction, followed by an evolutionary search of optimizing an expertise-inspired cost function that jointly considers the satisfaction of design rules. 
\item We conduct extensive evaluations on seven open-source designs against seven leading macro placers, using a mature backend implementation flow~\cite{ajayi2019toward} with sufficient timing optimizations, to validate performance under realistic industrial scenarios. Re$^{\text{2}}$MaP achieves the highest average rank on worst negative slack (WNS), total negative slack (TNS), and power. It is also comparable on routed wirelength, routing congestion, and runtime. Compared with the state-of-the-art metaheuristic-based method Hier-RTLMP~\cite{kahng2023hier}, Re$^{\text{2}}$MaP achieves up to 22.22\% (average 10.26\%) improvement in WNS and up to 97.91\% (average 33.97\%) improvement in TNS. Re$^{\text{2}}$MaP also demonstrates superior performance across varying core utilizations and target frequencies. A parameter tuning stage is further included, bringing significant timing and congestion improvement.
\end{itemize}

Compared to the conference version ReMaP~\cite{shi2025remap}, we propose a new macro grouping and PPA-aware clustering method that extends the original dataflow-only connections to a combined dataflow–wirelength matrix. We replace the fixed relocating order and greedy packing strategy with a robust, efficient packing tree-based relocating method that places groups of macros at each iteration. An efficient evolutionary search is followed to adjust the positions of macros within one group by optimizing an expertise-inspired cost function, explicitly targeting design rules. The evaluation flow is refined to support sufficient timing optimizations. To extend the evaluation scope, we include four more baseline methods (i.e., TritonMP~\cite{tritonmp}, RePlAce~\cite{cheng2018replace}, MaskPlace~\cite{lai2022maskplace}, and MaskRegulate~\cite{xue2024reinforcement}) and two more designs (i.e., \texttt{ariane133} and \texttt{swerv\_wrapper}) synthesized with the ASAP7 PDK. Compared to ReMaP, our new method demonstrates consistent average ranking improvements of 54.46\%, 47.90\%, 28.01\%, 19.14\%, and 11.32\% in WNS, TNS, power, design rule check (DRC) violations, and runtime, respectively. We also conduct experiments across different core utilizations and target frequencies.

\section{Proposed Re\texorpdfstring{\textsuperscript{2}}{2}MaP Method}

\definecolor{mycolor}{RGB}{255,217,102}
\begin{figure}[t]
\centering
\includegraphics[width=0.47\textwidth]{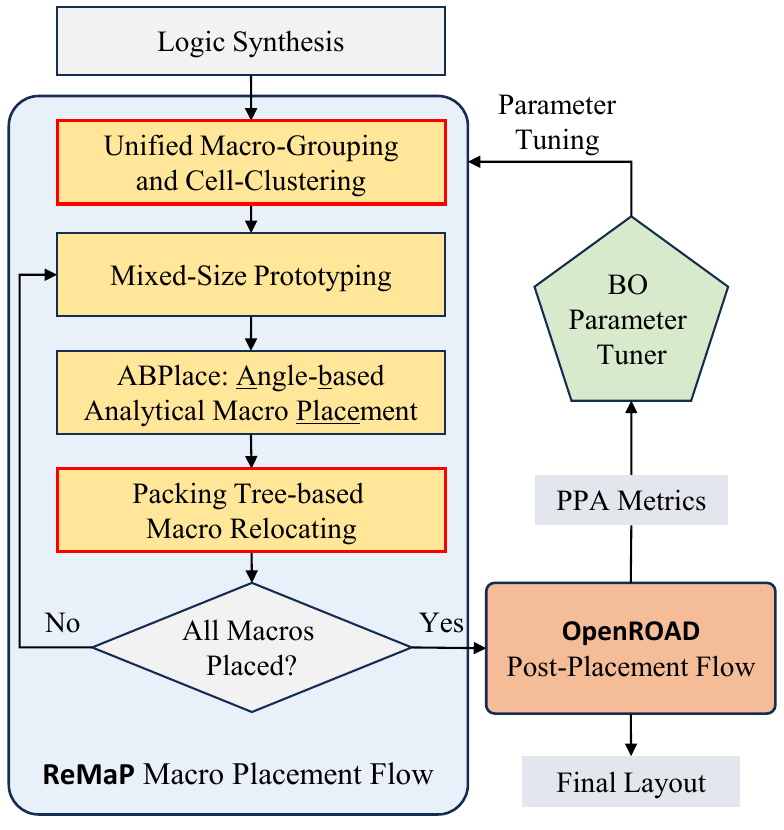}
\caption{Overview of the proposed Re$^{\text{2}}$MaP method. Components in \colorbox{myyellow}{yellow} are key steps of our method. Components bounded by \fcolorbox{red}{white}{red boxes} are key improvements over ReMaP~\cite{shi2025remap}.}
\label{fig:flow}
\end{figure}

We present Re$^{\text{2}}$MaP, a macro placement method that produces expert-quality layouts. As shown in Fig.~\ref{fig:flow}, the flow starts from a synthesized netlist. We first perform multi-level macro grouping and PPA-aware cell clustering. These steps produce macro groups, cell clusters, and a unified connection matrix that captures both direct connections and dataflow among macros and clusters. Next, we use DREAMPlace 4.1.0~\cite{lin2019dreamplace,chen2023stronger} to build a mixed-size placement prototype and obtain reference positions for each macro and cluster. Based on this prototype, we introduce ABPlace, an angle-based analytical method that optimizes macro positions on a predefined ellipse (or a circle for a square chip canvas). We then apply a packing tree-based relocating procedure that jointly optimizes the locations of macro groups and the macros within each group, using an expertise-inspired cost function that captures design constraints. In each iteration, we select a subset of macro groups and defer the remaining macros to the next iteration to improve the prototype’s accuracy. Finally, we perform full backend implementation with timing optimization, and evaluate power, performance, and area (PPA) using OpenROAD-flow-scripts~\cite{ajayi2019toward}. Additionally, a Bayesian optimization (BO)~\cite{frazier2018tutorial} parameter tuner is incorporated to adjust key parameters of Re$^{\text{2}}$MaP accordingly.

\subsection{Multi-Level Macro Grouping and PPA-Aware Clustering}\label{sec:2a}

\begin{figure}[t]
\centering
\includegraphics[width=0.46\textwidth]{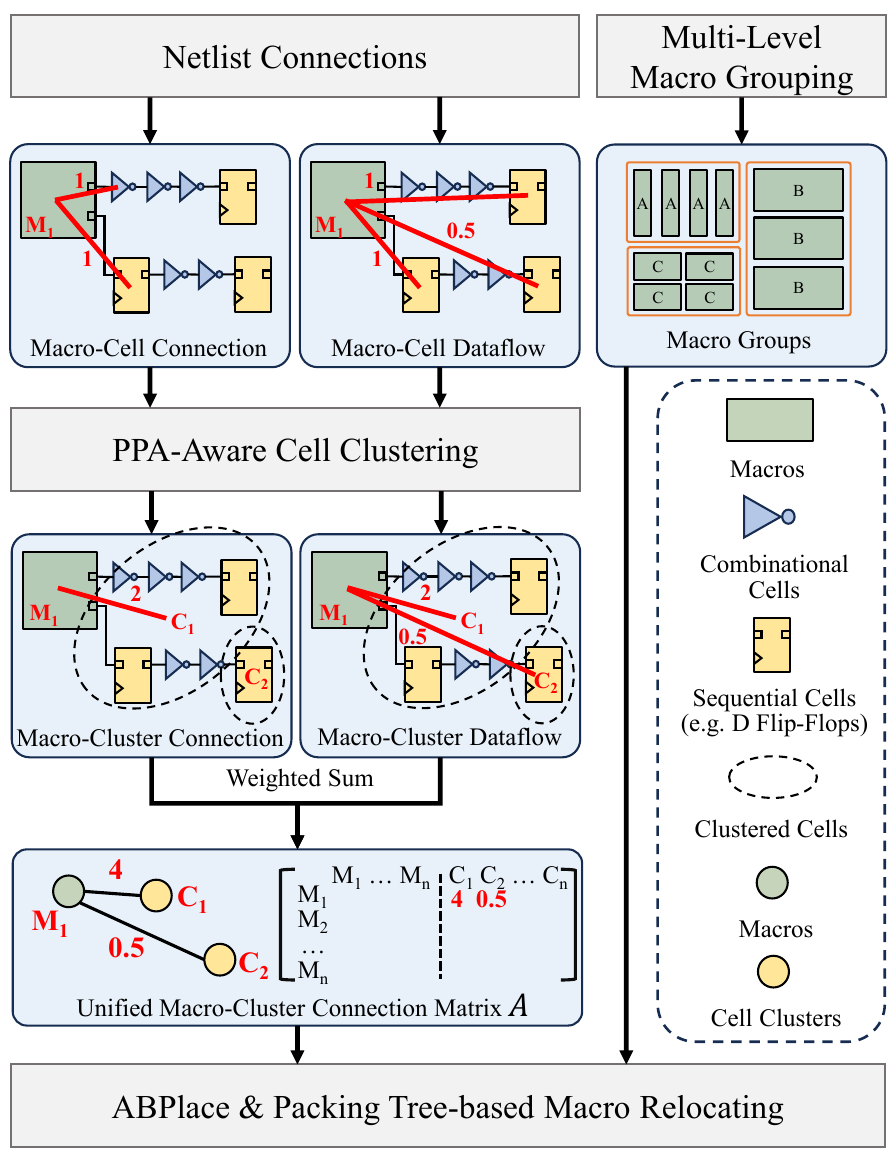}
\caption{Illustration of multi-level macro grouping and PPA-aware clustering pipeline.}
\label{fig:clustering}
\end{figure}

In this stage, we perform a pre-process that includes multi-level macro grouping and PPA-aware cell clustering. We obtain (1) grouped macros, (2) clustered cells, and (3) macro–macro, and macro-cell cluster connections from this stage. The overall pipeline is shown in Fig.~\ref{fig:clustering}.

\subsubsection{Macro Grouping}
In practice, engineers place macros using design hierarchy and RTL-level connection signatures~\cite{vidal2019rtl}. They also tend to tile macros of the same size and with similar connection signatures to improve regularity and reduce wirelength and congestion. To emulate this behavior and reveal RTL-level block connectivity, we adopt the multi-level auto-clustering from Hier-RTLMP~\cite{kahng2023hier}, considering both connection signatures and macro footprints. Concretely, we first cluster all macros by design hierarchy, then split out macros that differ in connection signature or footprint to form macro groups. This encourages regular tilings (e.g., 2$\times$4 arrays), leading to more regular placement and less dead space for subsequent cell placement. The macro grouping constraint will be considered during the relocating stage.

\subsubsection{PPA-Aware Cell Clustering}
Macro placement cannot ignore standard cells~\cite{zhao2024standard}. A common approach is to cluster standard cells to reduce problem size. Although Hier-RTLMP provides a clustered netlist for macros and cells, it does not incorporate PPA awareness during clustering and focuses primarily on macros, with limited optimization for standard cells. We therefore adopt the PPA-aware, cell-centered clustering method of \cite{kahng2024ppa}. Specifically, we weight nets by timing and power criticality, use a hierarchy-based multi-level clustering scheme, and apply a weighted-average Rent exponent criterion~\cite{cheon2002partition} to guide clustering optimization.

\subsubsection{Direct Connection and Dataflow Affinity}\label{sec:connection}
To guide macro placement, we build a connection matrix that captures both wirelength and dataflow affinity~\cite{vidal2019rtl,lin2021dataflow}. A detailed example is shown in Fig.~\ref{fig:clustering}. For wirelength extraction, we fully decompose each net by adding an undirected edge between every pair of instances inside the net. Using cell clustering, we aggregate edges between macros and between macros and clusters to derive macro-to-macro, and macro-to-cell-cluster wirelengths. For dataflow affinity extraction, we start from the original netlist that includes macros, combinational cells, and flip-flops. We remove all combinational cells (treating them as wires), leaving a directed graph whose nodes are macro pins and flip-flop pins. From each macro pin, we run a depth-first search with a maximum depth $D_{max}$ (set to three in our experiments). For every reachable sink pin $d$ (either a macro pin or a flip-flop pin) at depth $D \le D_{max}$, we add a directed edge from the source macro pin $s$ to $d$ with weight
\begin{align*}
w_{s\rightarrow d} = \frac{1}{2^D}.
\end{align*}
This yields weighted macro-to-macro and macro-to-flip-flop edges. And similar to direct connections, we aggregate these pin-level edges by the cell clustering to obtain macro–to-macro, and macro–to-cell-cluster dataflow affinities.

Finally, we normalize and summarize the wirelength and dataflow affinity terms to form a single unified macro-to-macro, and macro–to-cell-cluster connection matrix $A$.

\subsection{Recursive Mixed-Size Prototyping}\label{sec:2b}

\begin{figure*}[t]
\centering
\includegraphics[width=0.96\textwidth]{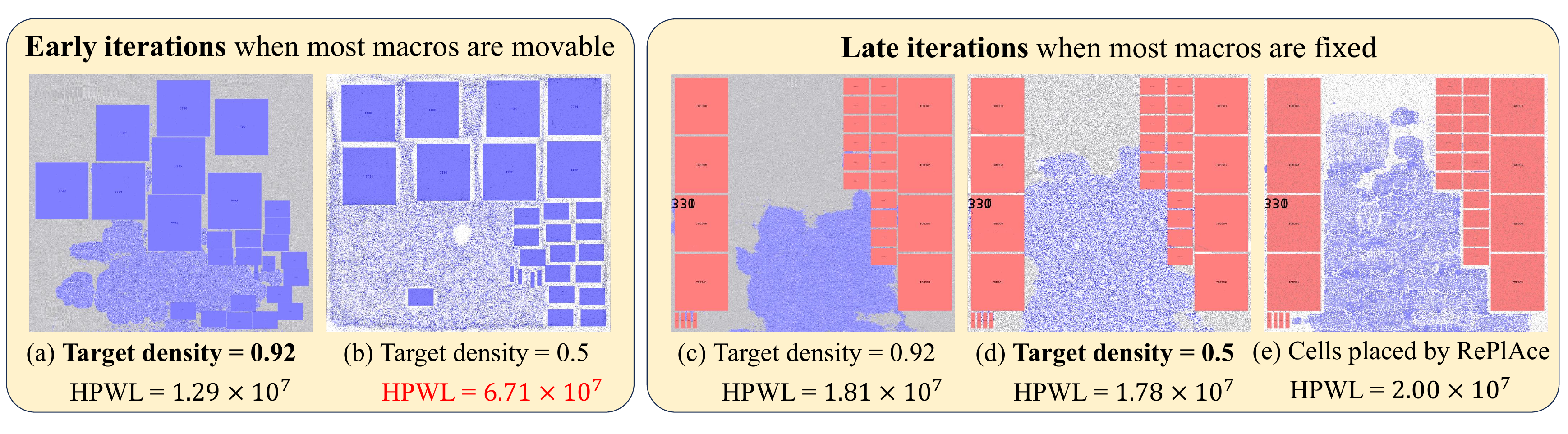}
\caption{Case study on \texttt{swerv\_wrapper} design demonstrates the effectiveness of the adaptive target density. Subfigures (a) and (b) show the impact of target density in the first iteration of mixed-size prototyping, where the placement of subfigure~(b) diverges and its HPWL is shown in \textcolor{red}{red}. Subfigures (c), (d), and (e) illustrate the cell placement after all macros have been fixed.}
\label{fig:target_density}
\end{figure*}

Macros do not directly determine design performance, but they interact with standard cells and thereby influence it~\cite{zhao2024standard}. In simplified scenarios that ignore power planning, signal integrity, and detailed routing, macro placement primarily affects global placement wirelength, which in turn impacts timing and power~\cite{mirhoseini2021graph,wang2024chipbench}.

A common strategy for incorporating standard cells into macro placement is to approximate standard cell clusters as soft blocks and co-optimize them with the macros~\cite{kahng2022rtl, kahng2023hier, lin2019regularity, lin2019routability}. However, estimating the area and aspect ratio of soft blocks is nontrivial~\cite{kahng2024ppa}, and the final cell locations can drift from the estimated cluster positions. As a result, mixed-size global placement is widely used to achieve competitive wirelength with modest runtime overhead~\cite{pu2024incremacro}. This mixed-size prototype provides an anchor for subsequent macro placement by serving as a reference that minimizes macro displacement~\cite{lin2022novel, chen2007mptrees}. Most prior work computes this prototype once, before all macro-optimization steps~\cite{lin2022novel, tritonmp, chen2007mptrees, lin2019novel}. However, such single-round prototype may become inaccurate as the process proceeds: Adjusting some macros can shift the ideal locations of other macros and of standard cells relative to the original prototype. Therefore, recomputing mixed-size prototypes after partial macro placement can provide higher-fidelity guidance.

We propose an iterative mixed-size prototyping method with an adaptive target density schedule. Each iteration proceeds as follows: (1) solve a mixed-size global placement over all movable macros and standard cells with an adjusted target density; (2) compute the center of gravity for each cell cluster; (3) select, optimize, and fix a subset of macros based on macro and cell cluster locations; (4) advance to the next iteration. In analytical placers (e.g., DREAMPlace~\cite{lin2019dreamplace}), target density is a key hyperparameter ranging from $(0, 1)$ that controls the final layout density. It starts high (e.g., $TD_{\text{init}} = 0.92$) in early iterations to improve convergence stability, as shown in Fig.~\ref{fig:target_density}~(a). When most macros are movable, mixed-size analytical placement exhibits notable randomness~\cite{yao2025evolution} and can diverge with a low target density, leading to large HPWL, as illustrated in Fig.~\ref{fig:target_density}~(b). We then gradually reduce the target density until the final iteration (e.g., $TD_{\text{finish}} = 0.5$), by which time most macros are fixed and convergence concerns are minimal, as shown in Fig.~\ref{fig:target_density}~(d). The low final density also matches that used in subsequent global cell placement, as shown in Fig.~\ref{fig:target_density}~(e), helping relieve routing congestion, and reducing legalization and buffering difficulty.

\subsection{ABPlace: Angle-based Analytical Macro Placement}\label{sec:ABPlace}

\begin{figure}[t]
\centering
\includegraphics[width=0.46\textwidth]{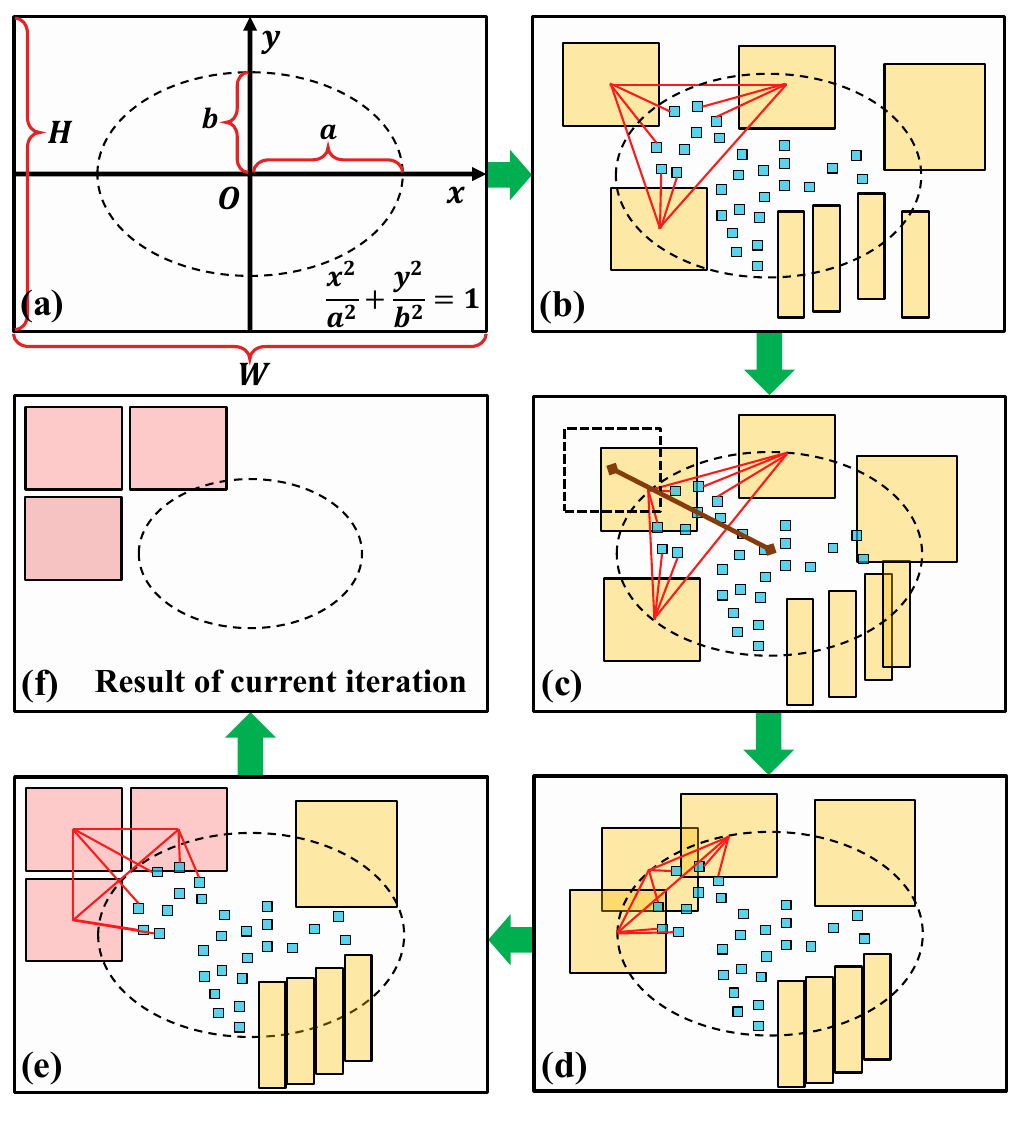}
\caption{Illustration of one Re$^\text{2}$MaP iteration. Subfigures (a)-(d) give the detailed steps from the ellipse construction to the result of ABPlace. Subfigures (e) and (f) give the illustration of macro relocating and the acquired results of the current iteration.}
\label{fig:abplace}
\end{figure}

After obtaining macro and cell-cluster positions from mixed-size prototyping, we apply ABPlace, an angle-based analytical macro placement on an ellipse. Such an optimization procedure primarily brings two advantages: It places macros onto an ellipse near their eventual periphery locations while considering connection cost and even distribution; it also reduces the two dimensional coordinate $(x, y)$ to an one dimensional polar angle $\theta$, which benefits convergence. ABPlace consists of three steps: ellipse construction, macro initialization, and analytical optimization.

\subsubsection{Ellipse Construction} We first construct an ellipse to further restrict the macros to its boundary, as shown in Fig.~\ref{fig:abplace}~(a). Let the canvas have width \( W \) and height \( H \). We use a 2D Cartesian coordinate system centered at the layout's geometric center; the \(x\)-axis is horizontal and the \(y\)-axis is vertical. The ellipse shares this center and is axis-aligned. Its semi-major axis \(a\) and semi-minor axis \(b\) are obtained by scaling the canvas half-dimensions by \(\beta_{\text{init}}\in(0,1)\) so the ellipse remains within the chip boundary, and then shrinking $a$ and $b$ per iteration by \(\gamma\in(0,1)\) so remaining macros are placed progressively nearer the center as outer regions become occupied, as shown in Fig.~\ref{fig:abplace}~(f) where the ellipse shrinks according to the crowdedness of the chip boundary areas. Consequently, at the \(k\)-th iteration we have the ellipse given by the following equations:
\begin{equation}
    \label{equation:ellipse_definition}
    \begin{cases}
        \displaystyle a = \beta_{\text{init}} \gamma^{k-1}\frac{W}{2}, \
        \displaystyle b = \beta_{\text{init}} \gamma^{k-1}\frac{H}{2},\\[8pt]
        \displaystyle \frac{x^2}{a^2} + \frac{y^2}{b^2} = 1,
    \end{cases}
\end{equation}
where \(x\) and \(y\) are coordinates in this chip-centered frame, and points \((x,y)\) satisfying the equation lie on the ellipse boundary. When \(W = H\), the ellipse reduces to a circle.

\subsubsection{Macro Initialization} After computing the prototyping locations, we construct the ellipse shown in Fig.~\ref{fig:abplace}~(b) and initialize placement by projecting every movable macro onto the ellipse boundary. Specifically, for each macro, we draw a ray from the ellipse center through the macro’s original center; the intersection of this ray with the ellipse boundary becomes the macro’s new position, as illustrated in Fig.~\ref{fig:abplace}~(c). Formally, given the original coordinates \((x_i^0, y_i^0)\) of macro \(i\), we define its mapped location \((x_i, y_i)\) as:
\begin{equation}
    \label{equation:theta_definition}
    \begin{cases}
        \displaystyle \theta_i = \arctan \frac{y_i^0}{x_i^0}, \\[8pt]
        \displaystyle x_i = a\cos\theta_i,\quad
        y_i = b\sin\theta_i,
    \end{cases}
\end{equation}
where \(\theta_i \in [0, 2\pi)\) is the polar angle of macro \(i\) with respect to the ellipse center.

\subsubsection{Analytical Optimization}\label{sec:analytical_optimization}
After mapping the macros onto the ellipse boundary, we optimize their placement using an angle-based formulation, where each macro’s position on the ellipse is uniquely determined by its angle. Let
\[
\bm{\theta} = [\theta_1, \theta_2, \ldots, \theta_n]^\top,\quad \theta_i \in [0, 2\pi),
\]
where $n$ denotes the total number of unplaced macros. \(\theta_i\) is the angle of macro $i$, and its corresponding coordinates are \(x_i = a\cos\theta_i\), \(y_i = b\sin\theta_i\), as given in Eq.~(\ref{equation:theta_definition}). Thus, optimizing macro locations reduces to optimizing \(\bm{\theta}\). To guide placement, we represent each cell cluster as a zero-area fixed anchor located at the centroid of its member cells. These anchors, the already placed macros, and the remaining unplaced macros together form the set of connection targets used to evaluate connection cost, which in turn determines the preferred location of each unplaced macro on the ellipse.

Our objective balances two goals: (1) minimizing the connection cost between macros and targets, and (2) encouraging a uniform, few-overlapping distribution of macros along the ellipse. The objective is formalized as:
\begin{equation}\label{eq:abplace}
    \min_{\bm{\theta}} \left(\sum_{i \in S_{\text{um}}}\sum_{j \in S_{\text{all}}}\text{dist}(i, j)A_{ij} + \lambda \sum_{i,j \in S_{\text{um}}, i \neq j}\text{overlap}(i,j)\right),
\end{equation}
where \(S_{\text{um}}\) is the set of unplaced macros, \(S_{\text{pm}}\) is the set of placed macros, \(S_{\text{cc}}\) is the set of cell clusters, \(S_{\text{all}} = S_{\text{um}} \cup S_{\text{pm}} \cup S_{\text{cc}}\) is the set of all macros and cell clusters, \(A\) is the connection strength matrix obtained from Sec.~\ref{sec:connection}, and \(A_{ij}\) denotes the $i$-th row and $j$-th column's entry of matrix \(A\). We use Euclidean distance for \(\text{dist}(\cdot,\cdot)\), and the overlap penalty between macros \(i\) and \(j\) is calculated as their actual overlapping area:
\begin{equation}
\begin{split}
\text{overlap}(i,j) = & \max\{(w_i+w_j)/2-|x_i-x_j|,0\} \\
& \times \max\{(h_i+h_j)/2-|y_i-y_j|,0\},
\end{split}
\end{equation}
where \(w_i\) and \(h_i\) are the width and height of macro \(i\). The overlap penalty uses $\max\{\cdot,\cdot\}$ and absolute value $|\cdot|$, which can cause unstable gradients. We implement the $\max\{\cdot,\cdot\}$ with \texttt{torch.clamp} with lower bound zero and replace \(|x|\) with the smooth surrogate \(\sqrt{x^2 + \epsilon^2} - \epsilon\). By setting \(\epsilon = 10^{-6}\), the surrogate is almost the same as \(|x|\) but stays differentiable at zero. The placement result of the analytical optimization is shown in Fig.~\ref{fig:abplace}~(d).

\subsection{Packing Tree-based Macro Relocating}\label{sec:macro_relocating}

\begin{algorithm}[t]
\caption{Macro Relocating}
\label{alg:relocating}

\textbf{Input}: $L_0$ \textcolor{commentcolor}{\textit{$\triangleright$ partially placed layout}}, $N_{\text{total}}$ \textcolor{commentcolor}{\textit{$\triangleright$ total number of evaluations}},
$N_{\text{eps}}$ \textcolor{commentcolor}{\textit{$\triangleright$ evaluations per slot}}, $N_{\text{pop}}$ \textcolor{commentcolor}{\textit{$\triangleright$ population size}}, $N_{\min}$ \textcolor{commentcolor}{\textit{$\triangleright$ minimum number of macros to place per iteration}}\\
\textbf{Output}: $L$ \textcolor{commentcolor}{\textit{$\triangleright$ layout with macros relocated}}
\begin{algorithmic}[1]
\State $Pref$ $\gets \mathrm{PreferenceCalculation}(L_0)$; \\ \Comment{calculate preference matrix according to layout $L_0$}
\State $L \gets L_0$; \Comment{initialize layout $L$ as $L_0$}
\State $N_{\text{placed}}$ $\gets 0$; \\ \Comment{number of macros that have been placed at this iteration}
\While{$N_{\text{placed}}$ $< N_{\min}$}
    \State $V$ $\gets \emptyset$; \Comment{valid packing set}
    \While{$V$ is empty}
        \State $i, j \gets \arg\max(Pref)$;\\ \Comment{i: macro group id, j: corner id}
        \State $V$ $\gets \mathrm{TryAssignment}(i, j, N_{\text{eps}})$; \\
        \Comment{return non-empty set if a valid solution exists}
        \If{$V$ is empty}
            \State $Pref_{ij} \gets -\infty$; \\
            \Comment{mask out infeasible assignment}
        \EndIf
    \EndWhile
    \State $P \gets \mathrm{CornerPackingSearching}(V, N_{\text{total}}, N_{\text{pop}})$;\\  \Comment{search for the best packing P using an evolutionary algorithm, taking $V$ as the initial population}
    \State $L \gets L.\mathrm{apply}(P)$; \\ \Comment{apply packing P to obtain the updated layout}
    \State $N_{\text{placed}} \gets N_{\text{placed}} + N_{i}$; \\ \Comment{$N_{i}$: number of macros in group i}
\EndWhile
\State \Return $L$
\end{algorithmic}
\end{algorithm}

In this section, we present a robust macro relocating method based on corner packing trees (see Sec.~\ref{sec:tree_representation}). The overall flow is shown in Alg.~\ref{alg:relocating}. The algorithm takes as input the placement produced by ABPlace (e.g., with cell clusters fixed and macros initially placed along an ellipse boundary), along with several hyperparameters. We first compute an assignment preference matrix $Pref$ that ranks macro groups and candidate packing corners (see Sec.~\ref{sec:preference_calculation}). We then attempt to implement the top-ranked assignment, checking for conflicts, while maintaining the packing structure with the best fitness for each feasible packing slot (see Sec.~\ref{sec:try_assignment}). If no feasible packing exists for the current choice, we reject it and move to the next preferred candidate. Once a macro group is successfully assigned to a corner, we search the corresponding corner tree for the best packing using an evolutionary algorithm~\cite{zhou2019evolutionary} (see Sec.~\ref{sec:corner_packing_searching}).

\subsubsection{Corner Packing Tree Representation} \label{sec:tree_representation}

\begin{figure}[t]
\centering
\includegraphics[width=0.48\textwidth]{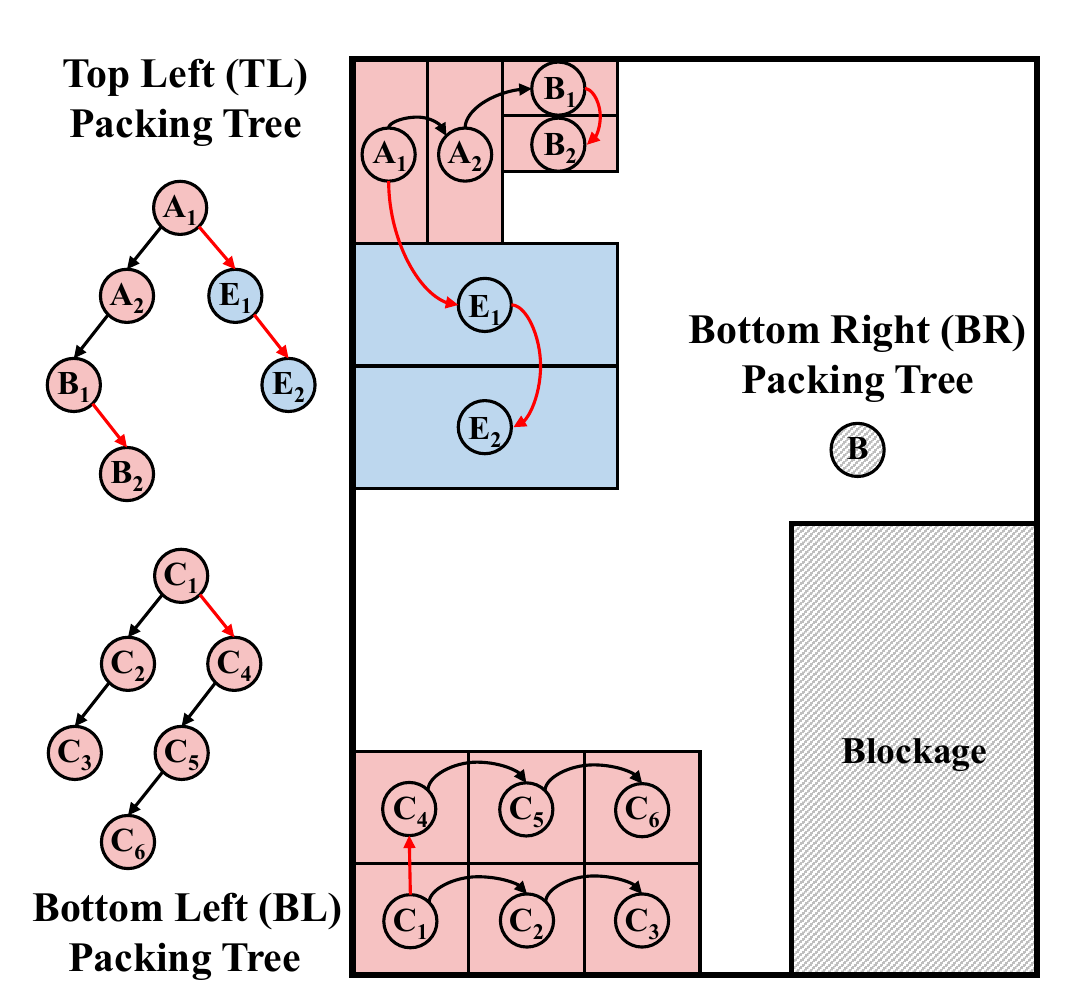}
\caption{Illustration of the packing tree representation. Black arrows indicate horizontal relationships, and \textcolor{red}{red} arrows indicate vertical relationships. Each macro block corresponds to a tree node. The \colorbox{mypink}{pink} blocks indicate already placed macros, and the \colorbox{myblue}{blue} blocks indicate the macro group placed at the current iteration. Specially, the BR packing tree consists of a placement blockage, which is also treated as an already placed macro.}
\label{fig:packing}
\end{figure}

We build four B*-trees~\cite{chang2000bstar} to pack macros into the four chip corners: bottom left (BL), bottom right (BR), top left (TL), and top right (TR), as shown in Fig.~\ref{fig:packing}. It is an intuitive way to implement a placement that pushes macros to the chip periphery. For each corner, we maintain a dedicated packing tree and its resulting placement. 

The detailed packing process is illustrated as follows. For a given corner, we start from its packing tree and place the root block directly at that corner. We then traverse the tree in pre-order. A left child is placed horizontally adjacent to its ancestor, and a right child is placed vertically adjacent to its ancestor. Because pre-order traversal packs horizontally before vertically, each macro can determine its $x$-coordinate by adding its ancestor’s width to the ancestor’s $x$-coordinate. For the $y$-coordinate, we maintain a contour that records the current packed height along the $x$-axis. After fixing the macro’s $x$-coordinate, we query the contour to obtain its $y$-coordinate, place the macro, and then update the contour by adding the macro’s height over its $x$-span.

To better explore the solution space, three mutation operators will be applied to the packing tree during optimization:
\begin{itemize}
    \item Swap: randomly select a node and exchange its left and right children.
    \item Rotate-L: randomly select a node and perform a left rotation.
    \item Rotate-R: randomly select a node and perform a right rotation.
\end{itemize}
We define a mutation sequence as repeatedly applying the above mutation operators. At each step, we select one operator with probability $p/3$ (for each of the three operators) and stop with probability $1-p$. This sequential scheme explores the search space more effectively than using a single operator, which often makes changes too small to escape local optima.

\subsubsection{Preference Calculation} \label{sec:preference_calculation}

Preference calculation is the first step of the relocating procedure because it determines which macro group to place and which corner to assign. Unlike methods that use a pre-defined placement order~\cite{lai2022maskplace,mirhoseini2021graph}, we jointly decide the placement order and the corner assignment to increase flexibility and improve performance.

We compute a preference matrix $Pref \in \mathbb{R}^{|\boldsymbol{G}|\times|\boldsymbol{C}|}$, where $|\boldsymbol{G}|$ is the number of unplaced macro groups and $|\boldsymbol{C}|$ is the number of corners (four in our setting). The $i$-th row and $j$-th column's entry of $Pref$ gives the preference for assigning macro group $i$ to corner $j$:
\begin{equation}
    Pref_{ij} = \alpha_1 \cdot A^{\text{MG}}_i - \left(\alpha_2 \cdot A^{\text{Util}}_j + \alpha_3 \cdot A^{\text{I/O}}_j + \alpha_4 \cdot \text{dist}(i,j)\right),
\end{equation}
where $A^{\text{MG}}_i$ is the total macro area of group $i$, $A^{\text{Util}}_j$ is the total area already utilized in corner $j$, $A^{\text{I/O}}_j$ is the area occupied by I/O ports in corner $j$, and $\text{dist}(i,j)$ is the Euclidean distance from the gravity center of group $i$ to corner $j$. Specially, if $A^{\text{I/O}}_j$ exceeds half of the total corner area, we directly ban all the assignments to this corner by setting $Pref_{\cdot j} = -\infty$. This rule comes from the observation that packing macros to the corner with too many IO ports will cause severe congestion for both I/O buffering and further routing. The weights $\alpha_{1-4}$ control the contribution of each term. 

\subsubsection{Try Assignment} \label{sec:try_assignment}

\begin{figure}[t]
\centering
\includegraphics[width=0.48\textwidth]{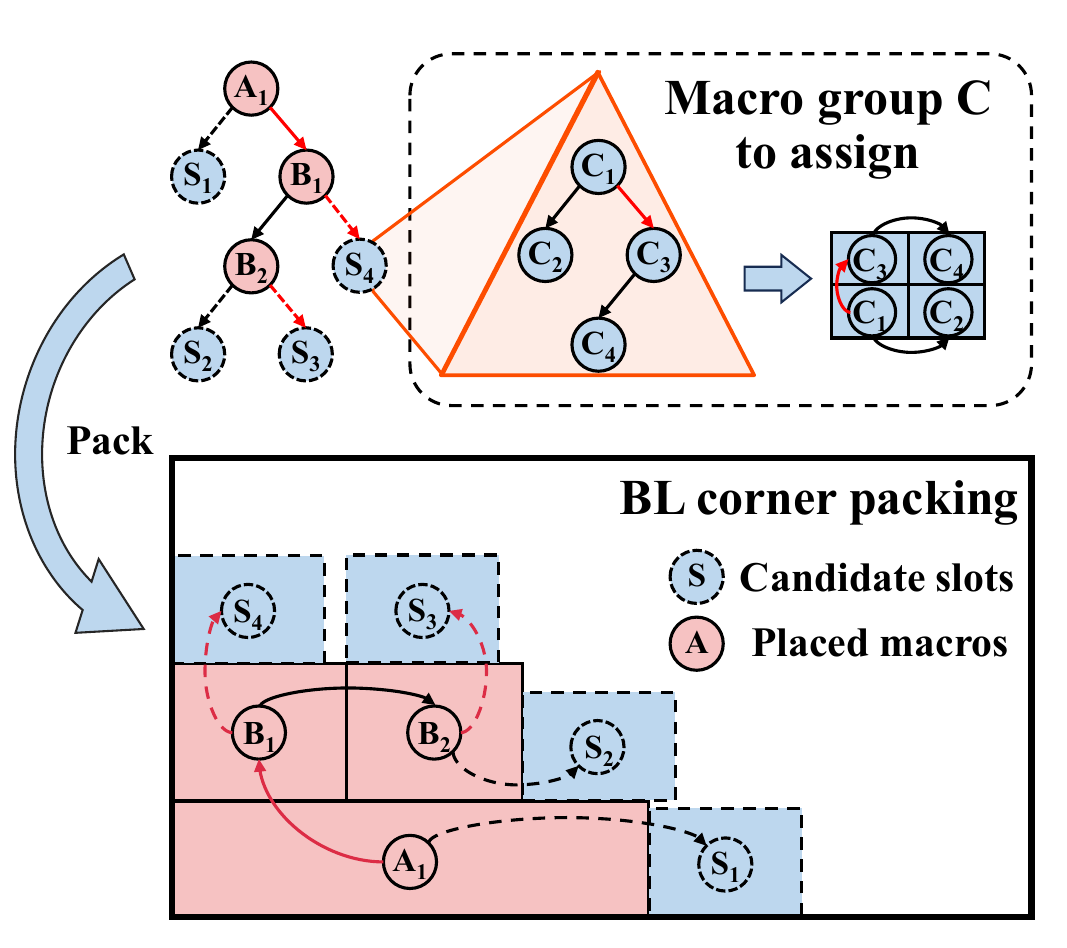}
\caption{Illustration of the procedure of TryAssignment. We try to pack macro group C onto the BL corner's packing tree. Each candidate slot is examined, and the best structure achieved so far is maintained for each slot.}
\label{fig:try_assignment}
\end{figure}

After calculating assignment preferences, we greedily choose the most preferred macro group and corner pair $(i,j)$ by $\arg\max Pref$, as shown in line 7 of Alg.~\ref{alg:relocating}. During the TryAssignment procedure, we search and validate all possible packing slots implied by this assignment. For each feasible slot, we retain the best packing found to seed the initial population for later evolutionary search.

The procedure is illustrated in Fig.~\ref{fig:try_assignment}. Suppose the preferred choice is group C, consisting of four identical macros $\text{C}_1$ through $\text{C}_4$, and the BL corner, which already has macros $\text{A}_1$, $\text{B}_1$, and $\text{B}_2$ packed with an existing packing tree. We aim to pack cluster C, a subtree consisting of four nodes, into the current tree. For a binary tree with $k$ nodes, there are $k+1$ empty leaf positions, and each such position is a candidate slot. Thus we have four candidate slots $\text{S}_1$ through $\text{S}_4$, marked by dotted circles. We attempt to attach subtree C at each candidate slot, starting from a randomly initialized subtree structure. The example highlighted by the orange triangle is intended for slot $\text{S}_4$. Because the four corner's trees are constructed independently and the initial subtree's shape can be irregular, some candidate slots may be infeasible. Three types of conflicts can occur: macro overlap with another corner’s packing tree, macro exceeding the chip boundary when too many macros are packed into the same corner, and violations that alter already placed macros in the existing tree. The last type arises from pre-order packing: Even without changing the structure of the already packed tree, previously placed nodes may be packed later than newly inserted nodes and thus lose their original positions.

For each slot, we randomly initialize the subtree, apply a mutation sequence defined in Sec.~\ref{sec:tree_representation}, evaluate the resulting packing, and retain the best structure found. $N_{\text{eps}}$ evaluations are performed per slot. If the packed layout exhibits any overlap, exceeds the placement boundary, or causes changes to already placed macros, we assign cost $+\infty$. Otherwise, the cost is given as follows:

\begin{equation}
\label{eq:fitness}
\begin{split}
        cost &= w_1 \cdot p_{\text{disp}} + w_2 \cdot p_{\text{conn}} + w_3 \cdot p_{\text{peri}} + w_4 \cdot p_{\text{group\_bb}} \\
    & \quad + w_5 \cdot p_{\text{corner\_bb}} + w_6 \cdot p_{\text{I/O}} + w_7 \cdot p_{\text{notch}},
\end{split}
\end{equation}
where each penalty is normalized across slots, and $w_1$ through $w_7$ are weights given by hyperparameters. The cost function consists of penalty terms that are inspired by design expertise, and target various design rules:
\begin{itemize}
    \item $p_{\text{disp}}$ is the displacement penalty, computed by summing the displacement of the group’s macros from their initial ellipse positions to the final locations.
    \item $p_{\text{conn}}$ is the connection penalty between macros in the current group and all placed macros and cell clusters,
    \begin{equation}
            p_{\text{conn}} = \sum_{i\in S_{\text{mg}}}\sum_{j\in S_{\text{pm}}\cup S_{\text{cc}}}\text{dist}(i,j)A_{ij},
    \end{equation}
    where $S_{\text{mg}}$ is the set of macros in the current group, $S_{\text{pm}}$ is the set of placed macros, and $S_{\text{cc}}$ is the set of cell clusters. We use Euclidean distance for $\text{dist}(\cdot,\cdot)$, and $A_{ij}$ is the $i,j$-th entry of connection matrix $A$ defined in Sec.~\ref{sec:connection}.
    \item $p_{\text{peri}}$ is the periphery cost, computed as the sum of each macro’s shortest distance from its edges to the chip boundary.
    \item $p_{\text{group\_bb}}$ is the penalty for the group-compactness constraint, defined as the area of the tight bounding box that encloses all macros belonging to the current group. Minimizing this term encourages each macro group to be compact and tile-like.
    \item $p_{\text{corner\_bb}}$ is the penalty for corner packing quality, defined as the sum of the areas of the tight bounding boxes for the four corner regions. Minimizing this term discourages macro sprawl within each corner and promotes tighter, more localized packing.
    \item $p_{\text{I/O}}$ penalizes occupation of I/O regions, computed as the total overlap area between macros and I/O regions.
    \item $p_{\text{notch}}$ penalizes notch area, i.e., small irregular regions between macro edges or between a macro and the chip boundary, that are unusable for analytic placement, computed as in Alg. 2 of RTL-MP~\cite{kahng2022rtl}.

\end{itemize}

After exhausting the evaluation budget, we discard infeasible slots with cost $+\infty$. All remaining feasible slots return their best packing solutions, forming a population for further searching and optimization. If no feasible slot exists, we reject the current macro group–corner assignment and consult the preference matrix for a new assignment, as in lines 9–10 of Alg.~\ref{alg:relocating}.

\subsubsection{Corner Packing Searching} \label{sec:corner_packing_searching}
We now have the best packing solution found so far for each slot. To further improve its quality, we run a longer search using an evolutionary algorithm~\cite{zhou2019evolutionary}. At each iteration, we perform tournament selection~\cite{Brindle1980Thesis} $N_{\text{pop}}$ times to obtain $N_{\text{pop}}$ parent solutions, apply a  mutation sequence to each parent to generate $N_{\text{pop}}$ offspring, and evaluate all offspring. We then combine the offspring with their parents, sort all candidates in increasing order of cost, and select the top-$N_{\text{pop}}$ solutions to form the next generation. We use a total evaluation budget of $N_{\text{total}}$. That is, the evolutionary search will be performed for $N_{\text{total}}/N_{\text{pop}}$ iterations, as each iteration needs to evaluate the newly generated $N_{\text{pop}}$ offspring solutions.

\section{Experimental Results}\label{sec:experiments}

\textbf{Environment:} We implement Re$^{\text{2}}$MaP in Python and evaluate it on a Linux server with two 64-core CPUs, one GPU with 48 GB memory, and 512 GB of RAM. We use seven RTL designs from the OpenROAD-flow-scripts\footnote{\url{https://github.com/The-OpenROAD-Project/OpenROAD-flow-scripts}} repository, synthesized with Nangate45 (NG45) PDK. For a more comprehensive analysis, we also synthesize two designs from MacroPlacement repository\footnote{\url{https://github.com/TILOS-AI-Institute/MacroPlacement}} with ASAP7 PDK. The statistics of our test cases are detailed in Table~\ref{table:statistics}. The released version of DREAMPlace 4.1.0\footnote{\url{https://github.com/limbo018/DREAMPlace/releases/tag/4.1.0}}~\cite{chen2023stronger} is used as our mixed-size placer for prototyping.

\begin{table}[ht]
\renewrobustcmd{\bfseries}{\fontseries{b}\selectfont}
\renewrobustcmd{\boldmath}{}
\newrobustcmd{\B}{\bfseries}
\caption{Detailed statistics of our test cases. We collect the number (\#) of macros, cells, and nets, while Freq. and Util. represent frequency and core utilization, respectively.}
\label{table:statistics}
\centering
\renewcommand{\arraystretch}{1.12}
\begin{tabular}{|c|c|c|c|c|c|}
\hline
\multirow{2}{*}{Design} & \multicolumn{1}{c|}{Macros} & \multicolumn{1}{c|}{Cells} & \multicolumn{1}{c|}{Nets} & \multicolumn{1}{c|}{Freq.} & \multirow{2}{*}{Util.} \\
& \multicolumn{1}{c|}{(\#)} & \multicolumn{1}{c|}{(\#)} & \multicolumn{1}{c|}{(\#)} & \multicolumn{1}{c|}{(MHz)} &  \\ \hline
\multicolumn{6}{|c|}{NG45} \\ \hline
\texttt{ariane133} & 132 & 196K & 207K & 333.3 & 0.36 \\ \hline
\texttt{ariane136} & 136 & 205K & 216K & 333.3 & 0.37 \\ \hline
\texttt{black\_parrot} & 24 & 342K & 369K & 500.0 & 0.50 \\ \hline
\texttt{bp\_be} & 10 & 61K & 67K & 555.6 & 0.48 \\ \hline
\texttt{bp\_fe} & 11 & 44K & 46K & 714.3 & 0.50 \\ \hline
\texttt{bp\_multi} & 26 & 180K & 191K & 1000.0 & 0.50 \\ \hline
\texttt{swerv\_wrapper} & 28 & 114K & 120K & 500.0 & 0.64 \\ \hline
\multicolumn{6}{|c|}{ASAP7} \\ \hline
\texttt{ariane133} & 133 & 163K & 180K & 250.0 & 0.25 \\ \hline
\texttt{swerv\_wrapper} & 28 & 105K & 114K & 400.0 & 0.15 \\ \hline
\end{tabular}
\end{table}

\textbf{Parameter settings:} For recursive mixed-size prototyping, we use an adaptive target density schedule with initial and final target densities of $TD_{\text{init}} = 0.92$ and $TD_{\text{finish}} = 0.5$. The per-iteration decay rate is $\sqrt[10]{TD_{\text{finish}} / TD_{\text{init}}}$. For ellipse construction, the initial scaling factor is $\beta_{\text{init}} = 0.9$, the final scaling factor is $\beta_{\text{finish}} = 0.5$, and the per-iteration shrinking factor is $\gamma = \sqrt[10]{\beta_{\text{finish}}/\beta_{\text{init}}}$. For analytical optimization on the ellipse, we set the Lagrange multiplier to $\lambda = 0.02$. The mutation probability factor is $p = 2/3$. For preference calculation, the weights for $\alpha_{1-4}$ are $5.0$, $0.5$, $4.0$, and $1.0$, respectively. For cost calculation, the weights for $w_{1-7}$ are $0.4$, $0.4$, $1.0$, $1.6$, $1.6$, $1.6$, and $1.0$, respectively. For evolutionary search, we set the total evaluation budget $N_{\text{total}}=100$, evaluation budget for each slot $N_{\text{eps}}=20$, and population size $N_{\text{pop}}=5$. The minimum number $N_{\text{min}}$ of macros to place at each iteration is $\left\lceil\frac{|S_{\text{um}}\cup S_{\text{pm}}|}{10}\right\rceil$, where $|S_{\text{um}}\cup S_{\text{pm}}|$ means the total number of macros.

\begin{figure}[t]
\centering
\includegraphics[width=0.48\textwidth]{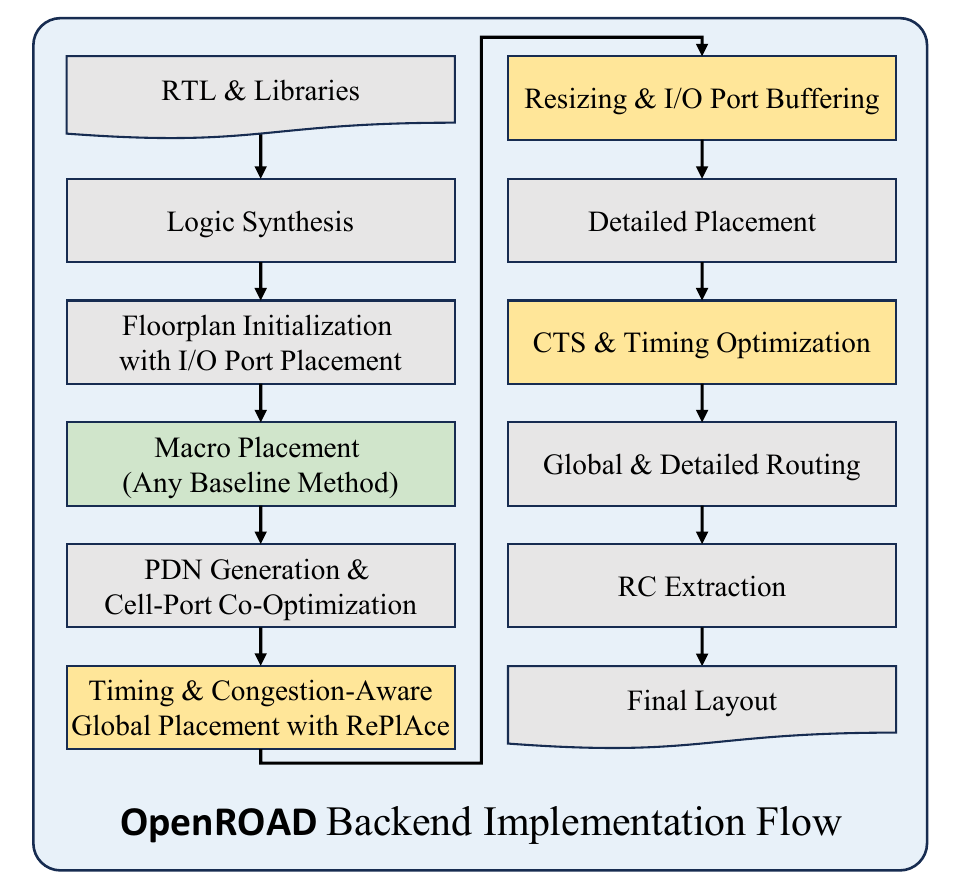}
\caption{OpenROAD backend implementation flow for PPA evaluation. The \colorbox{mygreen}{green} box highlights the macro placement stage performed by Re$^{\text{2}}$MaP or other baselines. The \colorbox{myyellow}{yellow} boxes highlight timing-optimization stages that target realistic implementation scenarios.}
\label{fig:openroad_flow}
\end{figure}

\textbf{Evaluation flow:} Fig.~\ref{fig:openroad_flow} outlines our pipeline using the OpenROAD released binary\footnote{\url{https://github.com/Precision-Innovations/OpenROAD/releases}}. We first synthesize RTL with Yosys\footnote{\url{https://github.com/YosysHQ/yosys}} to produce a gate-level netlist, then initialize the floorplan with outline $W$, $H$ and randomly placed I/Os. Then Re$^{\text{2}}$MaP is run to fix macro locations, followed by PDN construction, and co-optimize I/O with standard cells. Then RePlAce~\cite{cheng2018replace} is adopted to perform timing- and congestion-driven global placement, followed by buffering and sizing to mitigate design rule violations (DRVs) and detailed placement for legalization and wirelength reduction. Then We run clock tree synthesis (CTS) to build the clock network, followed by post-CTS buffering and sizing for timing optimization. After fixing all gates, we run global and detailed routing, reporting routed wirelength (rWL) and DRC count. Finally, we perform RC extraction and report final timing (WNS, TNS) and power.

\begin{table}[p]
\renewrobustcmd{\bfseries}{\fontseries{b}\selectfont}
\renewrobustcmd{\boldmath}{}
\newrobustcmd{\B}{\bfseries}
\caption{Post-route PPA performance on seven designs synthesized with NG45, comparing our \textit{Re$^\textit{2}$MaP} to \textit{TritonMP (TMP)}~\cite{tritonmp}, \textit{Hier-RTLMP (Hier-RTL.)}~\cite{kahng2023hier}, \textit{RePlAce}~\cite{cheng2018replace}, \textit{DREAMPlace} 4.1.0 \textit{(DMP)}~\cite{chen2023stronger}, \textit{MaskPlace (MaskPl.)}~\cite{lai2022maskplace}, \textit{MaskRegulate (MaskReg.)}~\cite{xue2024reinforcement}, and \textit{ReMaP}~\cite{shi2025remap}. Runtime is simplified as RT, and best results are in \textbf{bold}.}
\label{table:main}
\centering
\renewcommand{\arraystretch}{1.11}
\setlength{\tabcolsep}{5pt}
\begin{tabular}{|>{\begingroup\fontsize{7}{8}\selectfont}c<{\endgroup}|c|c|c|c|c|c|}
\hline
\multirow{2}{*}{\footnotesize Design} & {\multirow{2}{*}{Method}} & \multicolumn{1}{c|}{rWL} & \multicolumn{1}{c|}{WNS} & \multicolumn{1}{c|}{TNS} & \multicolumn{1}{c|}{Power} & \multicolumn{1}{c|}{RT} \\
 &  & \multicolumn{1}{c|}{(m)} & \multicolumn{1}{c|}{(ns)} & \multicolumn{1}{c|}{(ns)} & \multicolumn{1}{c|}{(mW)} & \multicolumn{1}{c|}{(s)} \\ \hline
\multirow{8}{*}{\texttt{ariane133}} 
 & \textit{TMP} & 7.55 & -1.18 & -3213 & 450 & 2159\\ 
 & \textit{Hier-RTL.} & 7.59 & -1.13 & -2892 & 448 & 90\\ 
 & \textit{RePlAce} & 7.27 & -1.13 & -2989 & 449 & 896\\
 & \textit{DMP} & 7.13 & -1.33 & -3640 & 454 & 36\\
 & \textit{MaskPl.} & 7.21 & -1.28 & -3760 & 451 & \textasciitilde 8h\\
 & \textit{MaskReg.} & 8.57 & -1.06 & -2877 & 449 & \textasciitilde 8h\\
 & \textit{ReMaP} & 7.49 & -1.12 & -2944 & \B 446 & 334\\
 & \textit{Re$^\textit{2}$MaP} & 7.94 & \B -1.05 & \B -2849 & 448 & 402\\ \hline
\multirow{8}{*}{\texttt{ariane136}} 
 & \textit{TMP} & 8.16 & -1.74 & -5132 & 506 & 5480\\ 
 & \textit{Hier-RTL.} & 7.85 & -1.19 & -3085 & 499 & 128\\ 
 & \textit{RePlAce} & 7.60 & -1.25 & -3296 & 508 & 919 \\
 & \textit{DMP} & 8.11 & -1.28 & -3517 & 518 & 47\\
 & \textit{MaskPl.} & 7.80 & -1.32 & -3713 & 498 & \textasciitilde 8h\\
 & \textit{MaskReg.} & 8.83 & -1.14 & -3003 & 505 & \textasciitilde 8h\\
 & \textit{ReMaP} & 7.84 & \B -1.06 & \B -2756 & 499 & 366\\
 & \textit{Re$^\textit{2}$MaP} & 8.02 & -1.09 & -2812 & \B 497 & 655\\ \hline
\multirow{8}{*}{\texttt{black\_parrot}} 
 & \textit{TMP} & 8.99 & -4.62 & -24 & 507 & 926\\ 
 & \textit{Hier-RTL.} & 9.81 & -4.63 & -918 & 498 & 71\\ 
 & \textit{RePlAce} & 9.79 & -4.83 & -160 & 496 & 942\\
 & \textit{DMP} & 9.50 & -4.71 & -129 & 495 & 45\\
 & \textit{MaskPl.} & 10.94 & -4.66 & -142 & 496 & \textasciitilde 8h\\
 & \textit{MaskReg.} & 10.17 & -4.76 & -56 & 550 & \textasciitilde 8h\\
 & \textit{ReMaP} & 8.86 & -4.74 & -46 & \B 493 & 240\\
 & \textit{Re$^\textit{2}$MaP} & 10.06 & \B -4.53 & \B -19 & 509 & 229\\ \hline
\multirow{8}{*}{\texttt{bp\_be}} 
 & \textit{TMP} & 2.70 & -1.11 & -222 & 202 & 817 \\ 
 & \textit{Hier-RTL.} & 3.06 & -1.24 & -382 & 206 & 13 \\ 
 & \textit{RePlAce} & 2.79 & -1.10 & -197 & 203 & 341\\
 & \textit{DMP} & 2.97 & -1.12 & -230 & 202 & 24\\
 & \textit{MaskPl.} & 2.63 & -1.06 & -226 & 203 & \textasciitilde 8h\\
 & \textit{MaskReg.} & 3.40 & -1.16 & -262 & 208 & \textasciitilde 8h\\
 & \textit{ReMaP} & 3.16 & -1.16 & -264 & 204 & 206\\
 & \textit{Re$^\textit{2}$MaP} & 2.69 & \B -1.06 & \B -193 & \B 201 & 64\\ \hline
\multirow{8}{*}{\texttt{bp\_fe}} 
 & \textit{TMP} & 1.99 & -0.49 & -60 & 209 & 744\\
 & \textit{Hier-RTL.} & 2.24 & -0.52 & -113 & 213 & 13\\ 
 & \textit{RePlAce} & 2.09 & -0.45 & -73 & 211 & 305\\
 & \textit{DMP} & 2.07 & -0.82 & -88 & 214 & 18\\
 & \textit{MaskPl.} & 2.21 & -0.53 & -90 & 215 & \textasciitilde 8h\\
 & \textit{MaskReg.} & 2.26 & -0.49 & -90 & 212 & \textasciitilde 8h\\
 & \textit{ReMaP} & 2.35 & \B -0.33 & \B -59 & 218 & 265\\
 & \textit{Re$^\textit{2}$MaP} & 1.81 & -0.47 & -62 & \B 208 & 80\\ \hline
\multirow{8}{*}{\texttt{bp\_multi}} 
 & \textit{TMP} & 5.06 & -5.33 & -2704 & 535 & 865\\ 
 & \textit{Hier-RTL.} & 5.05 & -5.77 & -2602 & 536 & 54\\ 
 & \textit{RePlAce} & 5.31 & -5.42 & \B -2452 & 542 & 575\\
 & \textit{DMP} & 5.21 & -5.76 & -2511 & 538 & 26\\
 & \textit{MaskPl.} & 6.51 & -5.58 & -4050 & 545 & \textasciitilde 8h\\
 & \textit{MaskReg.} & 6.86 & -5.80 & -3265 & 544 & \textasciitilde 8h\\
 & \textit{ReMaP} & 4.73 & -5.42 & -3486 & \B 530 & 325\\
 & \textit{Re$^\textit{2}$MaP} & 5.11 & \B -5.32 & -2683 & 536 & 172\\ \hline
\multirow{8}{*}{\texttt{swerv\_wrapper}} 
 & \textit{TMP} & 4.43 & -0.62 & -520 & 268 & 833\\ 
 & \textit{Hier-RTL.} & 5.50 & -0.72 & -649 & \B 267 & 31\\
 & \textit{RePlAce} & 5.37 & -0.71 & -635 & 273 & 595\\
 & \textit{DMP} & 4.91 & -0.73 & -698 & 270 & 22\\ 
 & \textit{MaskPl.} & 5.52 & -1.43 & -1410 & 273 & \textasciitilde 8h\\
 & \textit{MaskReg.} & 5.25 & -0.66 & -506 & 275 & \textasciitilde 8h\\
 & \textit{ReMaP} & 4.57 & -0.58 & -460 & 271 & 293\\
 & \textit{Re$^\textit{2}$MaP} & 4.51 & \B -0.56 & \B -401 & 269 & 146\\ \hline
\multicolumn{2}{|c|}{\multirow{2}{*}{Average Rank of \textit{ReMaP}}} & \multirow{2}{*}{3.86} & \multirow{2}{*}{3.14} & \multirow{2}{*}{3.57} & \multirow{2}{*}{3.57} & \multirow{2}{*}{3.71} \\
\multicolumn{2}{|c|}{} &  &  &  &  &  \\ \hline
\multicolumn{2}{|c|}{\multirow{2}{*}{Average Rank of \textit{Re$^\textit{2}$MaP} }} & \multirow{2}{*}{3.86} & \multirow{2}{*}{1.43} & \multirow{2}{*}{1.86} & \multirow{2}{*}{2.57} & \multirow{2}{*}{3.29} \\
\multicolumn{2}{|c|}{} &  &  &  &  &  \\ \hline
\end{tabular}
\end{table}

\begin{figure*}[t!]
  \centering
  \subfloat[\textit{TritonMP}~\cite{tritonmp}\label{fig:TritonMP}]{
    \includegraphics[width=0.24\textwidth]{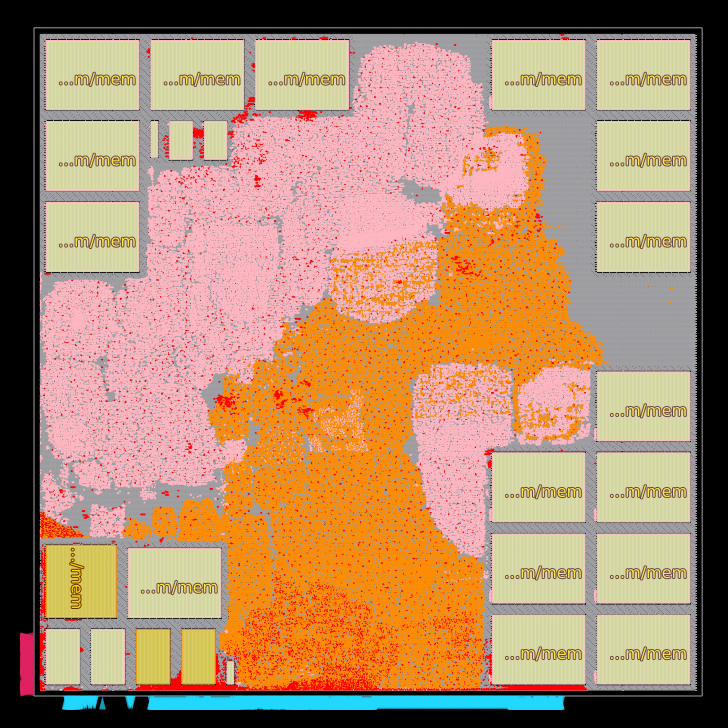}}\hfill
  \subfloat[\textit{Hier-RTLMP}~\cite{kahng2023hier}\label{fig:Hier-RTLMP}]{
    \includegraphics[width=0.24\textwidth]{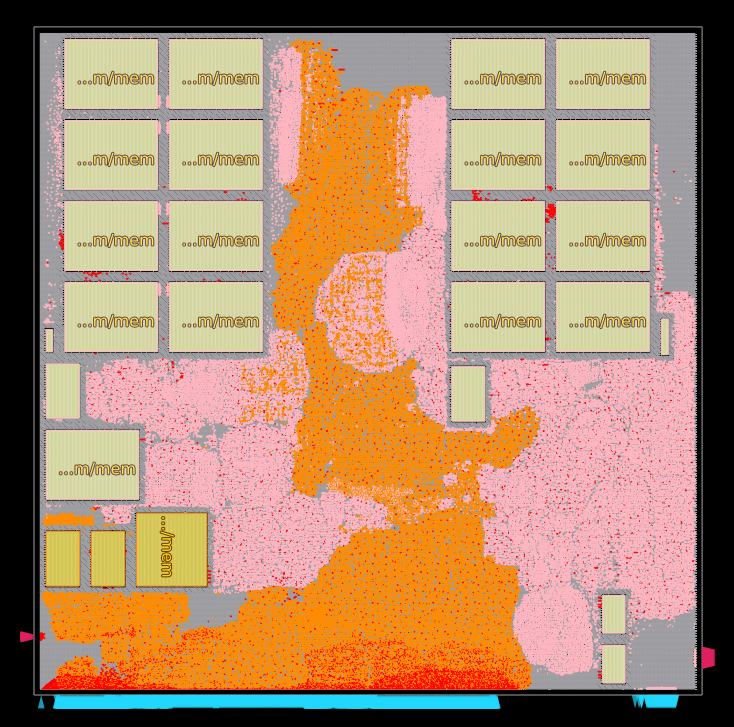}}\hfill
  \subfloat[\textit{RePlAce}~\cite{cheng2018replace}\label{fig:RePlAce}]{
    \includegraphics[width=0.24\textwidth]{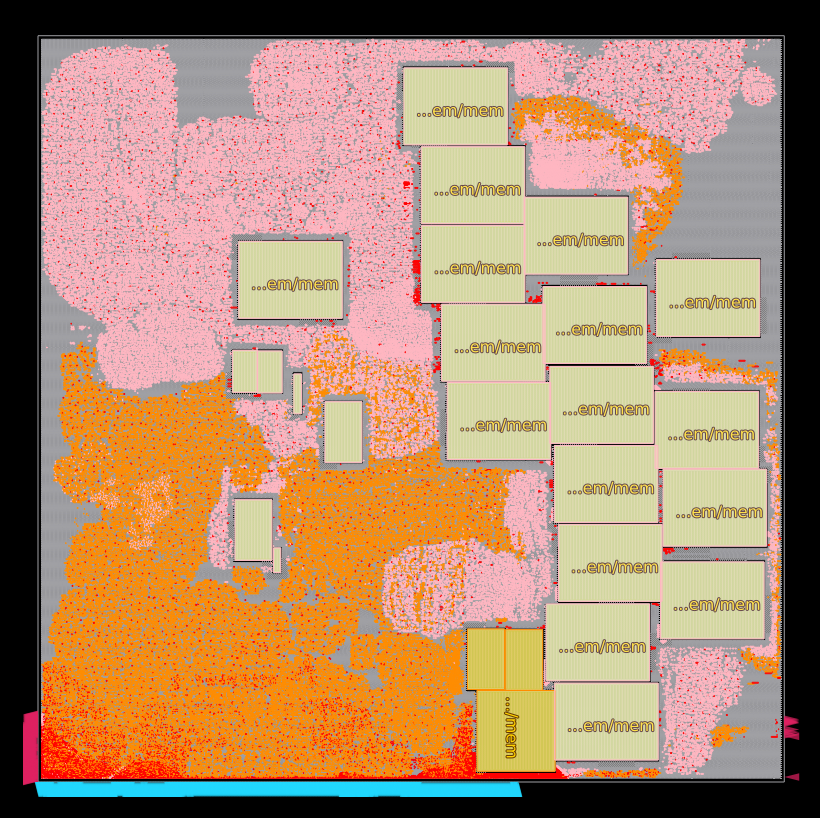}}\hfill
  \subfloat[\textit{DREAMPlace}~\cite{chen2023stronger}\label{fig:DREAMPlace}]{
    \includegraphics[width=0.24\textwidth]{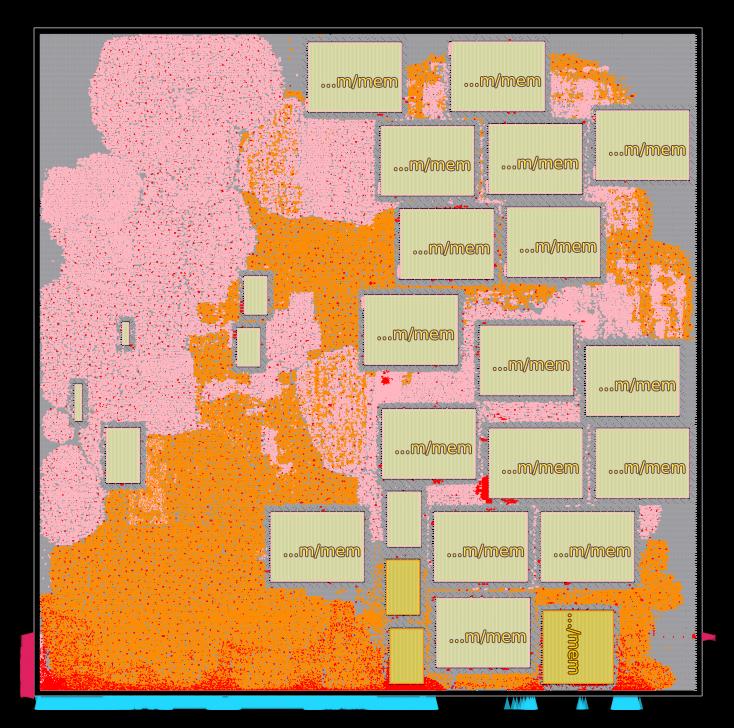}}\hfill
  \subfloat[\textit{MaskPlace}~\cite{lai2022maskplace}\label{fig:MaskPlace}]{
    \includegraphics[width=0.24\textwidth]{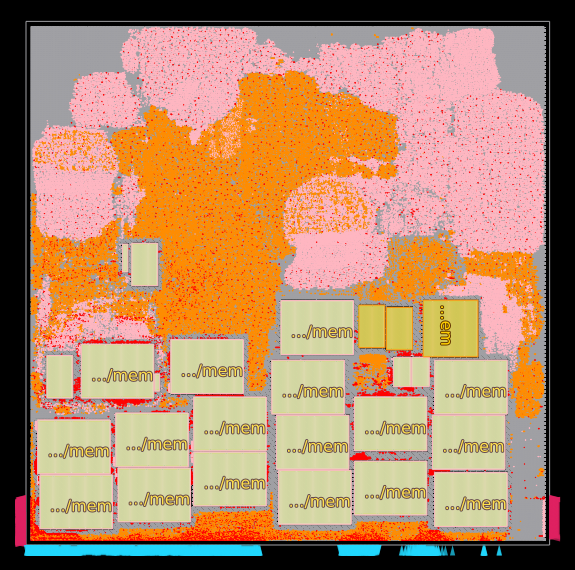}}\hfill
  \subfloat[\textit{MaskRegulate}~\cite{xue2024reinforcement}\label{fig:MaskRegulate}]{
    \includegraphics[width=0.24\textwidth]{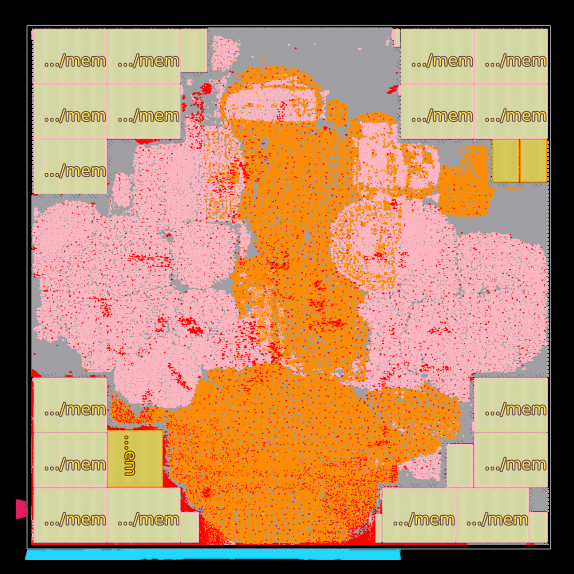}}\hfill
  \subfloat[\textit{ReMaP}~\cite{shi2025remap}\label{fig:ReMaP (old)}]{
    \includegraphics[width=0.24\textwidth]{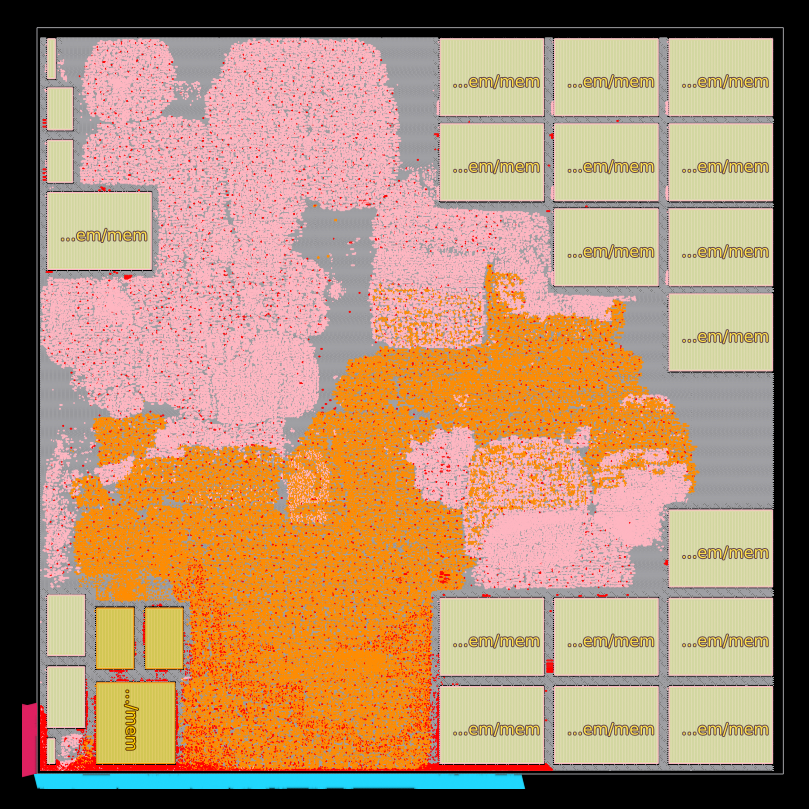}}\hfill
  \subfloat[\textit{Re$^\textit{2}$MaP}\label{fig:ReMaP (new)}]{
    \includegraphics[width=0.24\textwidth]{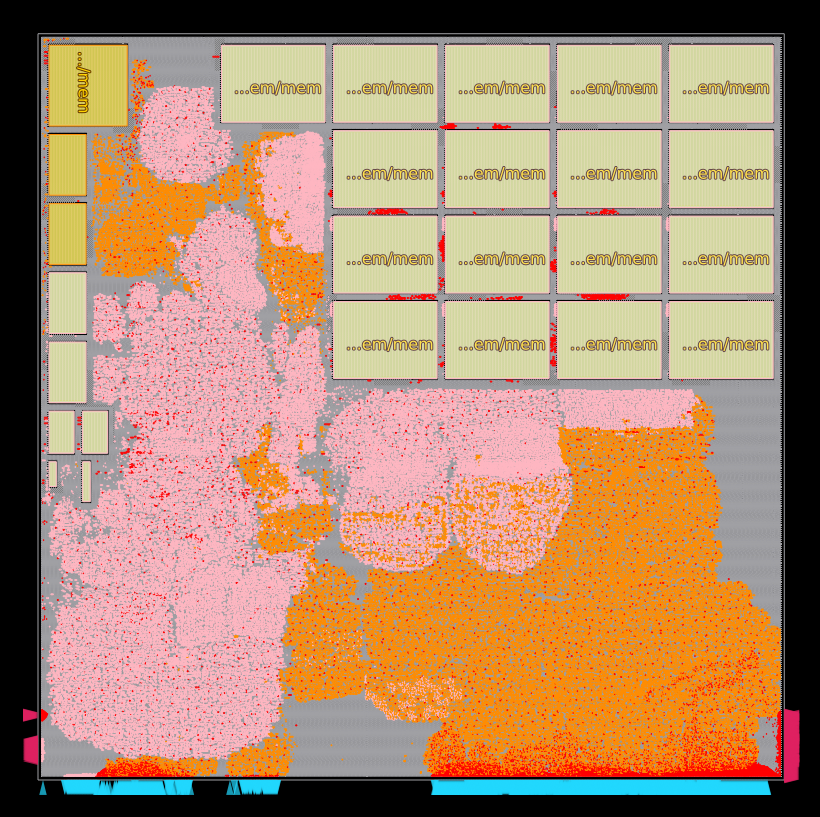}}    
  \caption{Placement visualization of design \texttt{bp\_multi}. \colorbox{myyellow}{Yellow} blocks represent macros. \colorbox{mypink}{Pink} and \colorbox{myorange}{orange} areas represent standard cells belonging to different hierarchies. \textcolor{red}{Red} dots represent buffers inserted. The \textcolor{pinblue}{blue} and \textcolor{red}{red} dotted lines lying on the chip's bottom boundary are stacked triangles representing I/O ports.}
  \label{fig:placements}
\vspace{-10pt}
\end{figure*}

\subsection{Main Results}

\textbf{Baselines:} We compare our \textit{Re$^\textit{2}$MaP} against seven baselines in Table~\ref{table:main}, including traditional metaheuristic-based placers (\textit{TritonMP}~\cite{tritonmp}, \textit{Hier-RTLMP}~\cite{kahng2023hier}); analytical mixed-size placers (\textit{RePlAce}~\cite{cheng2018replace}, \textit{DREAMPlace}~4.1.0~\cite{chen2023stronger}); learning-based methods (\textit{MaskPlace}~\cite{lai2022maskplace}, \textit{MaskRegulate}~\cite{xue2024reinforcement}); and the conference version \textit{ReMaP}~\cite{shi2025remap}. \textit{TritonMP} and \textit{Hier-RTLMP} are integrated into OpenROAD and run directly in our evaluation flow. For \textit{RePlAce}, we use the OpenROAD implementation\footnote{\url{https://github.com/The-OpenROAD-Project/RePlAce}} and enable mixed-size placement with the \texttt{timing\_driven} and \texttt{congestion\_driven} options. Since OpenROAD does not provide a stand alone macro legalizer, we just report the PPA of macro placements with minor overlaps produced by \textit{RePlAce}. For \textit{DREAMPlace}, we run mixed-size placement with \texttt{macro\_place\_flag} and \texttt{use\_bb}, followed by an integrated advanced macro legalization procedure. For \textit{MaskPlace}\footnote{\url{https://github.com/laiyao1/maskplace}} and \textit{MaskRegulate}\footnote{\url{https://github.com/lamda-bbo/macro-regulator}}, we use their open-source code, train each case from scratch for about eight hours, and select the macro placement with the best reward. For \textit{ReMaP}, we adopt the released code\footnote{\url{https://github.com/lamda-bbo/DAC25-ReMaP}} to obtain its results.

\textbf{Performance analysis:} \textit{Re$^\textit{2}$MaP} delivers the best timing in most cases, ranking first on $5/7$ designs for WNS and $4/7$ for TNS, with average ranks of 1.43 (WNS) and 1.86 (TNS) across all eight methods (see the last row of Table~\ref{table:main}). Our method also achieves an average rank of 2.57 in power, likely because better macro organization reduces the number of inserted buffers. For routing congestion (measured by DRC count), our approach yields relatively low congestion on most designs, with an average rank of 3.00, notably better than \textit{RePlAce}~\cite{cheng2018replace} and \textit{MaskPlace}~\cite{lai2022maskplace}, which do not explicitly enforce periphery placement or I/O region keepout. \textit{TritonMP}~\cite{tritonmp} performs best on DRC due to its corner-based packing and regularity-driven optimization; however, it overlooks dataflow connections and macro grouping constraints, leading to poor timing performance. In runtime, we consistently outperform \textit{TritonMP} which requires time-consuming SA search, \textit{RePlAce} which relies on CPU to perform analytical optimization, and \textit{MaskPlace} or \textit{MaskRegulate}~\cite{xue2024reinforcement} requiring long training time. Our total runtime remains reasonable: about 10 minutes for the largest design $\texttt{ariane136}$ and about 1 minute for small designs. Runtime analysis is further detailed in Sec.~\ref{sec:runtime}. For rWL, our rank is not significantly higher because rWL is influenced by multiple backend steps and has notable variability. In fact, \textit{TritonMP} ranks first in rWL with an average rank of 3.29, while \textit{DREAMPlace}~\cite{chen2023stronger}, \textit{ReMaP}~\cite{shi2025remap}, and \textit{Re$^\textit{2}$MaP} tie for second with 3.86.

\textbf{Comparison with \textit{ReMaP}}: The newly proposed \textit{Re$^\textit{2}$MaP} shows consistent superiority in all key PPA metrics, as illustrated in the last two rows of Table~\ref{table:main}, with an average ranking improvements of 54.46\%, 47.90\%, 28.01\%, 19.14\%, and 11.32\% in WNS, TNS, power, DRC count, and runtime, respectively. The timing performance gains are the most pronounced, driven by the new clustering scheme and the packing tree-based relocating optimization that facilitates effective searching. The added I/O region keepout reserves space for port buffering and routing, improving both timing and congestion. Despite the more advanced clustering method and the tree search procedure, our overall efficiency still improves, as detailed in Sec.~\ref{sec:runtime}.

\textbf{Visualization:} Fig.~\ref{fig:placements} shows final layouts from eight methods on \texttt{bp\_multi}. \textit{TritonMP} packs macros near the four corners, creating large continuous cell regions and achieving the best routability. \textit{Hier-RTLMP} groups macros into tiles and considers I/O region keepout, but often introduces notches and irregular macro placements. \textit{RePlAce} and \textit{DREAMPlace} yield similar macro distributions; however, \textit{RePlAce} tends to place macros more densely, sometimes with overlaps. Such irregularly placed macros introduce dead space and complicate buffering and routing, and they also prevent a robust PDN since the cells surrounded by macros are hardly reachable. Both methods largely ignore periphery placement and explicit I/O region keepout, which hurts routing and buffering, leading to uncompetitive final routed wirelength and congestion. \textit{MaskPlace} optimizes macro-to-macro connectivity, which produces crowded macro clusters and overlooks I/O regions, resulting in weak performance on most cases. \textit{MaskRegulate} uses a periphery-aware reinforcement learning agent to push macros toward corners but lacks macro halos and explicit I/O region keepout. It delivers better timing and DRC than \textit{MaskPlace}, yet remains inferior to traditional methods. Compared to \textit{ReMaP}, \textit{Re$^\textit{2}$MaP} introduces I/O region keepout that reserves port areas for buffering and routing. It also places macros group-by-group, yielding more regular placements and improved timing.

\subsection{Testing on Designs Using ASAP7 PDK}
We synthesize \texttt{ariane133} and \texttt{swerv\_wrapper} using the ASAP7 PDK and compare our method with \textit{TritonMP}~\cite{tritonmp} and \textit{Hier-RTLMP}~\cite{kahng2023hier}. Shown in Table~\ref{table:asap7}, the metric WS denotes worst slack, which is introduced since there are no negative-slack endpoints in \texttt{swerv\_wrapper}. \textit{Hier-RTLMP} fails on \texttt{ariane133}, likely due to version incompatibility. Our \textit{Re$^\textit{2}$MaP} achieves the best performance on timing and DRC clean on both designs. In particular, our method shows largest positive slack on \texttt{swerv\_wrapper}, indicating that \textit{Re$^\textit{2}$MaP} can increase the target frequency without introducing timing violations.

\begin{table}[ht]
\renewrobustcmd{\bfseries}{\fontseries{b}\selectfont}
\renewrobustcmd{\boldmath}{}
\newrobustcmd{\B}{\bfseries}
\caption{PPA results using designs synthesized with ASAP7 PDK, comparing to \textit{TritonMP (TMP)}~\cite{tritonmp} and \textit{Hier-RTLMP (Hier-RTL.)}~\cite{kahng2023hier},}\label{table:asap7}
\centering
\renewcommand{\arraystretch}{1.12}
\begin{tabular}{|c|c|c|c|c|}
\hline
\multirow{2}{*}{Design} & \multirow{2}{*}{Method} & \multicolumn{1}{c|}{WS} & \multicolumn{1}{c|}{TNS} & \multicolumn{1}{c|}{DRC} \\ 
 &  & \multicolumn{1}{c|}{(ns)} & \multicolumn{1}{c|}{(ns)} & \multicolumn{1}{c|}{(\#)} \\ \hline
 \multirow{3}{*}{\texttt{ariane133}} 
 & \textit{TMP} & -1.72 & -2559 & 0 \\ 
 & \textit{Hier-RTL.} & - & - & - \\  
 & \textit{Re$^\textit{2}$MaP} & \B -1.70 & \B -2534 & 0 \\ \hline
\multirow{3}{*}{\texttt{swerv\_wrapper}} 
 & \textit{TMP} & 0.34 & 0 & 0 \\ 
 & \textit{Hier-RTL.} & 0.69 & 0 & 0 \\  
 & \textit{Re$^\textit{2}$MaP} & \B 0.74 & 0 & 0 \\ \hline
\end{tabular}
\end{table}

\subsection{Testing on High-Utilization Designs}

Since most of our benchmarks have about 50\% utilization, we manually shrink the chip outline to assess performance at a higher utilization. The results are summarized in Table~\ref{table:area_curve}. Starting from \texttt{swerv\_wrapper} with 64\% utilization, we further reduce the canvas to reach 70\%, 75\%, and 80\% utilization—near the practical limit for very crowded designs. If we include macro halos into the utilization calculation, the area utilization will surpass 90\%. In these over-crowded settings, our method maintains strong timing and achieves the best DRC results, owing to a balanced use of periphery placement and I/O region keepout. At 80\% utilization, \textit{Hier-RTLMP} fails due to insufficient area during its optimization, and \textit{TritonMP} also exhibits poor timing and severe congestion.

\begin{table}[ht]
\renewrobustcmd{\bfseries}{\fontseries{b}\selectfont}
\renewrobustcmd{\boldmath}{}
\newrobustcmd{\B}{\bfseries}
\caption{Testing on high-utilization designs, comparing to \textit{TritonMP (TMP)}~\cite{tritonmp} and \textit{Hier-RTLMP (Hier-RTL.)}~\cite{kahng2023hier}, to validate the robustness of our method on congested layouts.}\label{table:area_curve}
\centering
\renewcommand{\arraystretch}{1.12}
\begin{tabular}{|c|c|c|c|c|}
\hline
\multirow{2}{*}{Design} & \multirow{2}{*}{Method} & \multicolumn{1}{c|}{WNS} & \multicolumn{1}{c|}{TNS} & \multicolumn{1}{c|}{DRC} \\ 
 &  & \multicolumn{1}{c|}{(ns)} & \multicolumn{1}{c|}{(ns)} & \multicolumn{1}{c|}{(\#)} \\ \hline
 \multirow{3}{*}{\makecell{\texttt{swerv\_wrapper}\\ \texttt{64 Util.}}}
 & \textit{TMP} & -0.62 & -520 & 1494   \\ 
 & \textit{Hier-RTL.} & -0.72 & -649 & 4841 \\  
 & \textit{Re$^\textit{2}$MaP} & \B -0.56 & \B -401 & \B 789 \\\hline
 \multirow{3}{*}{\makecell{\texttt{swerv\_wrapper}\\ \texttt{70 Util.}}}
 & \textit{TMP} & -0.63 & -695 & 3795 \\ 
 & \textit{Hier-RTL.} & \B -0.49 & \B -400 & 4196 \\  
 & \textit{Re$^\textit{2}$MaP} & -0.71 & -457 & \B 740 \\\hline
 \multirow{3}{*}{\makecell{\texttt{swerv\_wrapper}\\ \texttt{75 Util.}}}
 & \textit{TMP} & -0.76 & \B -624 & 7930 \\ 
 & \textit{Hier-RTL.} & \B -0.73 & -677 & 11905 \\  
 & \textit{Re$^\textit{2}$MaP} & \B -0.73 & -637 & \B 889 \\\hline
 \multirow{3}{*}{\makecell{\texttt{swerv\_wrapper}\\ \texttt{80 Util.}}}
 & \textit{TMP} & -1.94 & -702 & 639956 \\ 
 & \textit{Hier-RTL.} & - & - & - \\  
 & \textit{Re$^\textit{2}$MaP} & \B -0.72 & \B -664 & \B 3228 \\\hline
\end{tabular}
\end{table}

\subsection{Testing under High and Low Speed Settings}
Modern designs target a wide range of clock frequencies and require different levels of timing optimization effort. We therefore evaluate whether our method produces placements that work well at both high and low speeds. As shown in Table~\ref{table:timing_curve}, we sweep the target frequency from 1000 MHz down to 400 MHz, re-running the entire flow each time (synthesis, placement, timing optimization, and final evaluation). Across this sweep, our \textit{Re$^\textit{2}$MaP} outperforms \textit{TritonMP} and \textit{Hier-RTLMP} on both WNS and TNS, indicating better compatibility with timing optimization. Although our DRC counts are not lower than \textit{TritonMP}, the resulting congestion is much milder than \textit{Hier-RTLMP} and can be resolved quickly with a few additional detailed routing optimization rounds.

\begin{table}[h]
\renewrobustcmd{\bfseries}{\fontseries{b}\selectfont}
\renewrobustcmd{\boldmath}{}
\newrobustcmd{\B}{\bfseries}
\caption{Testing under high and low speed settings, comparing to \textit{TritonMP (TMP)}~\cite{tritonmp} and \textit{Hier-RTLMP (Hier-RTL.)}~\cite{kahng2023hier}, to validate the reliability of our method under both aggressive and relaxed speed targets.}\label{table:timing_curve}
\centering
\renewcommand{\arraystretch}{1.12}
\begin{tabular}{|c|c|c|c|c|}
\hline
\multirow{2}{*}{Design} & \multirow{2}{*}{Method} & \multicolumn{1}{c|}{WNS} & \multicolumn{1}{c|}{TNS} & \multicolumn{1}{c|}{DRC} \\ 
 &  & \multicolumn{1}{c|}{(ns)} & \multicolumn{1}{c|}{(ns)} & \multicolumn{1}{c|}{(\#)} \\ \hline

 \multirow{3}{*}{\makecell{\texttt{bp\_be}\\ \texttt{1000 MHz}}}
 & \textit{TMP} & -1.96 & -4134 & \B 19 \\
 & \textit{Hier-RTL.} & -2.09 & -3461 & 2549 \\
 & \textit{Re$^\textit{2}$MaP} & \B -1.85 & \B -2930 & 144 \\\hline

 \multirow{3}{*}{\makecell{\texttt{bp\_be}\\ \texttt{666.7 MHz}}}
 & \textit{TMP} & -1.45 & -957.46 & \B 10 \\
 & \textit{Hier-RTL.} & -1.56 & -1138 & 2721 \\
 & \textit{Re$^\textit{2}$MaP} & \B -1.42 & \B -948 & 137 \\\hline

 \multirow{3}{*}{\makecell{\texttt{bp\_be}\\ \texttt{555.6 MHz}}}
 & \textit{TMP} & -1.10 & -222 & \B 15 \\
 & \textit{Hier-RTL.} & -1.24 & -382 & 2400 \\
 & \textit{Re$^\textit{2}$MaP} & \B -1.05 & \B -193 & 153 \\\hline

 \multirow{3}{*}{\makecell{\texttt{bp\_be}\\ \texttt{500 MHz}}}
 & \textit{TMP} & -0.94 & -142 & \B 15 \\
 & \textit{Hier-RTL.} & -1.06 & -171 & 2395 \\
 & \textit{Re$^\textit{2}$MaP} & \B -0.84 & \B -111 & 98 \\\hline

 \multirow{3}{*}{\makecell{\texttt{bp\_be}\\ \texttt{400 MHz}}}
 & \textit{TMP} & -0.47 & -47 & \B 27 \\
 & \textit{Hier-RTL.} & -0.55 & -55 & 2691 \\
 & \textit{Re$^\textit{2}$MaP} & \B -0.36 & \B -34 & 40 \\\hline
\end{tabular}
\end{table}

\subsection{Ablation Study}\label{sec:ablation}

We conduct an ablation study on the \texttt{swerv\_wrapper} design to isolate the effect of each component of our algorithm. The results are summarized in Table~\ref{table:ablation}. The \textit{w/o ABPlace} variant maps macros to the ellipse only and skips analytical optimization. The \textit{w/o Ellipse} variant disables ellipse construction and instead performs packing tree-based macro relocating guided directly by mixed-size prototyping. The \textit{w/o Macro Group} variant removes macro grouping from the relocating procedure. The \textit{w/o Ellipse Shrinking} variant keeps the ellipse at its initial size and does not shrink it across iterations. The \textit{w/o Dynamic Density} variant keeps the target density at its initial value $TD_{\text{init}}$ and does not adjust it across iterations. The \textit{w/o I/O Keepout} variant ignores I/O region keepout during macro relocating, which makes legalization failed due to congestion near ports. \textit{ReMaP} is the conference version that adopts Louvain clustering~\cite{fogaca2020superiority} and a greedy periphery-guided relocating method. Across all variants, the default \textit{Re$^\textit{2}$MaP} achieves the best timing while maintaining comparable routing congestion.

\begin{table}[ht]
\renewrobustcmd{\bfseries}{\fontseries{b}\selectfont}
\renewrobustcmd{\boldmath}{}
\newrobustcmd{\B}{\bfseries}
\caption{Ablation study to validate the contribution of each component in our method.}\label{table:ablation}
\centering
\renewcommand{\arraystretch}{1.12}
\begin{tabular}{|c|c|c|c|c|}
\hline
\multirow{2}{*}{Design} & \multirow{2}{*}{Variations} & \multicolumn{1}{c|}{WNS} & \multicolumn{1}{c|}{TNS} & \multicolumn{1}{c|}{DRC} \\ 
 &  & \multicolumn{1}{c|}{(ns)} & \multicolumn{1}{c|}{(ns)} & \multicolumn{1}{c|}{(\#)} \\ \hline
 \multirow{8}{*}{\texttt{swerv\_wrapper}} 
 & \textit{w/o ABPlace} & -0.65 & -645 & \B 662 \\ 
 & \textit{w/o Ellipse} & -0.71 & -477 & 735 \\  
 & \textit{w/o Macro Group} & -0.78 & -479 & 1440 \\ 
 & \textit{w/o Ellipse Shrinking} & -0.69 & -436 & 773 \\
 & \textit{w/o Dynamic Density} & -0.64 & -415 & 1097 \\
 & \textit{w/o I/O Keepout} & - & - & - \\
 & \textit{ReMaP}~\cite{shi2025remap} & -0.58 & -460 & 1473 \\
 & \textit{Re$^\textit{2}$MaP} & \B -0.56 & \B -401 & 789 \\\hline
\end{tabular}
\end{table}

\subsection{Parameter Tuning}\label{sec:hpo}

To further exploit the capability of our algorithm and obtain better placements, we apply Bayesian optimization (BO)~\cite{frazier2018tutorial} to tune key hyperparameters. We consider eight continuous parameters in the range $(0,1)$: the Lagrangian multiplier $\lambda$ used in ABPlace analytical optimization in Sec.~\ref{sec:analytical_optimization}, and the seven weights $w_{1-7}$ used during relocating cost computation in Sec.~\ref{sec:try_assignment}. We formulate the parameter tuning problem as:
\begin{equation}
    \min_{\lambda,w_{1-7}} \left(\frac{\text{WNS}}{\text{WNS}_\text{orig}} + \frac{\text{TNS}}{\text{TNS}_\text{orig}} + \frac{\text{DRC}}{\text{DRC}_\text{orig}}\right),
\end{equation}
where $\text{WNS}_\text{orig}$, $\text{TNS}_\text{orig}$, and $\text{DRC}_\text{orig}$ denote the metrics obtained with the default parameters, and WNS, TNS, and DRC are the metrics obtained with the tuned parameters.

Table~\ref{table:hpo} reports the tuning results. For each design, we run 50 iterations of BO. In particular, for \texttt{bp\_fe}, tuning yields an additional improvement of 44.7\% in WNS, 56.5\% in TNS, and achieves 100\% DRC clean. Because the full backend flow includes extensive timing optimization, each evaluation is time-consuming. Therefore, we use two small designs for the search, and each evaluation takes about 30 minutes. Completing 50 iterations requires roughly 24 hours per design.

\begin{table}[ht]
\renewrobustcmd{\bfseries}{\fontseries{b}\selectfont}
\renewrobustcmd{\boldmath}{}
\newrobustcmd{\B}{\bfseries}
\caption{Results of automatic parameter tuning. Eight key parameters are chosen to be optimized by a 50-rounds of BO.}\label{table:hpo}
\centering
\renewcommand{\arraystretch}{1.15}
\begin{tabular}{|c|c|c|c|c|}
\hline
\multirow{2}{*}{Design} & \multirow{2}{*}{Method} & \multicolumn{1}{c|}{WNS} & \multicolumn{1}{c|}{TNS} & \multicolumn{1}{c|}{DRC} \\ 
 &  & \multicolumn{1}{c|}{(ns)} & \multicolumn{1}{c|}{(ns)} & \multicolumn{1}{c|}{(\#)} \\ \hline
 \multirow{2}{*}{\texttt{bp\_be}} 
 & \textit{Default Params.} & -1.06 & -193 & 153\\ 
 & \textit{50-rounds of BO}  & \B -1.04 & \B -187 & \B 83 \\ \hline
 \multirow{2}{*}{\texttt{bp\_fe}} 
 & \textit{Default Params.} & -0.47 & -62 & 97 \\ 
 & \textit{50-rounds of BO}  & \B -0.26 & \B -27 & \B 0 \\ \hline
\end{tabular}
\end{table}

\subsection{Runtime Analysis}\label{sec:runtime}

Table~\ref{table:rumtime} reports a detailed runtime breakdown of the \textit{Re$^\textit{2}$MaP} method and compares it with \textit{ReMaP}. We observe a consistent reduction in the total number of iterations, which directly reduces the number of prototyping steps, ABPlace rounds, and relocating search rounds. As a result, \textit{Re$^\textit{2}$MaP} spends less time on DREAMPlace input parsing (I/O), DREAMPlace gradient computation (DMP-Gradient), and ABPlace, leading to a significant total runtime reduction on \texttt{black\_parrot}, \texttt{bp\_be}, \texttt{bp\_fe}, \texttt{bp\_multi}, and \texttt{swerv\_wrapper}. The reduction in the number of iterations stems from macro grouping and a more effective relocating strategy. In both \textit{ReMaP} and \textit{Re$^\textit{2}$MaP}, we set the same budget $N_{\text{min}} = \left\lceil\frac{|S_{\text{um}}\cup S_{\text{pm}}|}{10}\right\rceil$ to bound the total number of macros placed at each iteration, targeting a total iteration number of 10. For \textit{ReMaP} the budget is checked once a macro is placed, while for \textit{Re$^\textit{2}$MaP} the macros are decided group by group, and the budget is checked only after a group is completed. Thus \textit{Re$^\textit{2}$MaP} often place more macros than the budget per iteration. Besides, \textit{ReMaP} may fail to relocate since the feasible placement regions are decided according to the whitespace between the ellipse and the chip boundary. If the ellipse shrinks too slow, more iterations are needed to facilitate feasible relocating. For \textit{Re$^\textit{2}$MaP}, however, feasible locations are guaranteed in every iteration due to the packing tree formulation.

\begin{figure}[t]
\centering
\includegraphics[width=0.47\textwidth]
{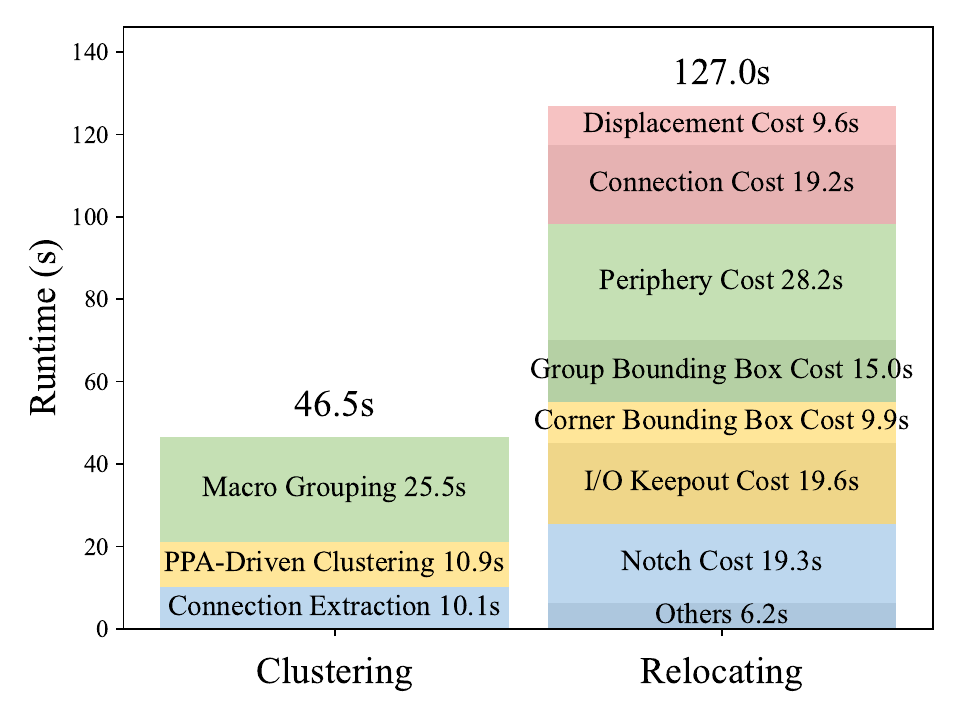}
\caption{Runtime breakdown of Clustering and Relocating steps of \textit{Re$^\textit{2}$MaP} on design \texttt{ariane133}.}
\label{fig:runtime}
\vspace{-0.65\baselineskip}
\end{figure}

For clustering, although the new approach includes multi-level optimization procedures, they run efficiently and scale better than the prior Louvain clustering. The detailed runtime of each component is shown in the left column of Fig.~\ref{fig:runtime}. For relocating, shown in the right column of Fig.~\ref{fig:runtime}, most of the runtime is spent on cost computation, with the components listed individually. The relocating overhead is particularly high for \texttt{ariane133} and \texttt{ariane136} because, as the number of macros increases, the number of candidate slots to evaluate for each corner packing tree also increases, which raises the cost computation time.

\begin{table*}[!ht]
\renewrobustcmd{\bfseries}{\fontseries{b}\selectfont}
\renewrobustcmd{\boldmath}{}
\newrobustcmd{\B}{\bfseries}
\caption{Detailed runtime decomposition, showing an efficiency gain compared to \textit{ReMaP}~\cite{shi2025remap} .}\label{table:rumtime}
\centering
\renewcommand{\arraystretch}{1.15}
\setlength{\tabcolsep}{6pt}
\begin{tabular}{|c|c|c|c|c|c|c|c|c|c|}
\hline
\multirow{2}{*}{Design} & \multirow{2}{*}{Method} & \multicolumn{1}{c|}{Iterations} & \multicolumn{1}{c|}{Clustering} & \multicolumn{1}{c|}{I/O} & \multicolumn{1}{c|}{DMP-Gradient} & \multicolumn{1}{c|}{ABPlace} & \multicolumn{1}{c|}{Relocating} & \multicolumn{1}{c|}{Others} & \multicolumn{1}{c|}{Total}\\
 &  & \multicolumn{1}{c|}{(\#)} & \multicolumn{1}{c|}{(s)} & \multicolumn{1}{c|}{(s)} & \multicolumn{1}{c|}{(s)} & \multicolumn{1}{c|}{(s)} & \multicolumn{1}{c|}{(s)} & \multicolumn{1}{c|}{(s)} & \multicolumn{1}{c|}{(s)}\\ \hline
\multirow{2}{*}{\texttt{ariane133}} 
& \textit{ReMaP} & 13 & 24.3 & 141.3 & 98.8 & 55.9 & 3.5 & 10.7 & 334.5 \\ 
& \textit{Re$^\textit{2}$MaP} & 10 & 46.5 & 105.4 & 92.7 & 23.7 & 127.0 & 6.7 & 402.1 \\ \hline
\multirow{2}{*}{\texttt{ariane136}} 
& \textit{ReMaP} & 14 & 27.8 & 157.1 & 104.8 & 60.4 & 3.6 & 11.8 & 365.5 \\
& \textit{Re$^\textit{2}$MaP} & 9 & 59.3 & 129.4 & 85.0 & 18.2 & 361.5 & 1.4 & 654.8 \\ \hline
\multirow{2}{*}{\texttt{black\_parrot}} 
& \textit{ReMaP} & 6 & 23.5 & 112.4 & 68.5 & 27.8 & 0.8 & 7.3 & 240.3 \\
& \textit{Re$^\textit{2}$MaP} & 5 & 25.5 & 117.7 & 69.0 & 9.9 & 5.3 & 1.5 & 228.8 \\ \hline
\multirow{2}{*}{\texttt{bp\_be}} 
& \textit{ReMaP} & 10 & 8.8 & 26.4 & 128.1 & 38.0 & 0.5 & 4.6 & 206.2 \\ 
& \textit{Re$^\textit{2}$MaP} & 4 & 6.7 & 12.4 & 32.6 & 7.8 & 3.6 & 1.4 & 64.4 \\ \hline
\multirow{2}{*}{\texttt{bp\_fe}} 
& \textit{ReMaP} & 13 & 9.1 & 26.0 & 174.0 & 49.4 & 1.1 & 5.6 & 265.3 \\
& \textit{Re$^\textit{2}$MaP} & 4 & 5.9 & 11.2 & 49.6 & 7.7 & 4.8 & 0.7 & 79.8 \\ \hline
\multirow{2}{*}{\texttt{bp\_multi}} 
& \textit{ReMaP} & 13 & 23.6 & 110.7 & 125.0 & 54.2 & 1.6 & 10.2 & 325.3 \\
& \textit{Re$^\textit{2}$MaP} & 5 & 20.2 & 62.3 & 65.6 & 10.6 & 11.9 & 1.1 & 171.7 \\ \hline
\multirow{2}{*}{\texttt{swerv\_wrapper}} 
& \textit{ReMaP} & 11 & 14.4 & 60.0 & 166.6 & 44.0 & 1.0 & 6.6 & 292.7 \\
& \textit{Re$^\textit{2}$MaP} & 5 & 12.3 & 47.4 & 58.1 & 10.4 & 16.8 & 0.9 & 145.8 \\ \hline
\end{tabular}
\end{table*}

\section{Conclusion}\label{sec:conclusion}

In this paper, we present Re$^{\text{2}}$MaP, a macro placement method that combines recursive analytical prototyping with an expertise-inspired metaheuristic. The flow integrates multi-level macro grouping and PPA-aware cell clustering into a unified connection matrix capturing both wirelength and dataflow. Based on mixed-size prototype, ABPlace optimizes macro angles on an ellipse to achieve periphery-aware, connectivity-driven placement. A packing tree–based relocating procedure jointly optimizes group locations and intra-group macro positions, guided by an evolutionary search over a comprehensive design constraint-aware cost. These stages are performed recursively for more precise prototypings. Extensive experiments have demonstrated the effectiveness of our method, compared with a wide variety of leading macro placers. We have open-sourced the whole method to promote reproducible research. We hope the proposed method Re$^{\text{2}}$MaP can provide a practical path to expert-quality, scalable macro placement.

\bibliographystyle{IEEEtran}
\bibliography{main}

\end{document}